\documentclass[aps,prb,reprint,superscriptaddress,amsmath,amssymb,showpacs]{revtex4-1}

\usepackage{graphicx}
\usepackage{color}
\usepackage{amsmath,amssymb,mathrsfs,ulem}
\usepackage[caption=false]{subfig}
\usepackage{bbm}
\usepackage[dvipsnames]{xcolor}
\usepackage{tikz}
\usepackage{pgfplots}
\usetikzlibrary{calc,arrows,decorations.pathmorphing,decorations.markings,backgrounds,positioning,fit,petri,shapes,intersections,patterns}
\newcommand{\ket}[1]{\left|#1\right\rangle}
\newcommand{\bra}[1]{\left\langle#1\right|}
\newcommand{\expect}[1]{\langle #1\rangle}
\newcommand{\lket}[1]{|#1)}
\newcommand{\lbra}[1]{(#1|}
\newcommand{\dunderline}[1]{\underline{\underline{#1}}}
\newcommand{\td}[2]{{\mathrm{d} #1 \over \mathrm{d} #2}}
\newcommand{\com}[2]{[\,#1\,,\,#2\,]}
\newcommand{\anticom}[2]{\lbrace\,#1\,,\,#2\,\rbrace}
\newcommand{\abs}[1]{\vert #1\vert}
\newcommand{\orderof}[1]{\mathcal{O}\left(#1\right)}

\DeclareMathOperator{\sgn}{sgn}
\DeclareMathOperator{\tr}{Tr}
\DeclareMathOperator{\real}{Re}
\DeclareMathOperator{\imag}{Im}

\tikzstyle{node1}=[draw,circle,ultra thick,inner sep=0pt,minimum size=7]
\tikzstyle{node3}=[draw,strike out, ultra thick]
\tikzstyle{node4}=[draw,cross out, ultra thick]
\tikzstyle{line1}=[line width=2pt]
\tikzstyle{selfenergy}=[draw,ellipse,inner sep=0pt,minimum size=15]
\tikzstyle{derivative}=[draw,circle,ultra thick,fill=black,inner sep=0pt,minimum size=3]
\tikzstyle{amputated}=[dashed,line width=2pt]
\tikzset{
    singlescale/.style={draw=black,double,line width=1pt, postaction={decorate},
        decoration={markings,mark=at position .5 with {\node[node3]{};}}},
}

\date{\today}                  

\begin{document}

\title{RG transport theory for open quantum systems: Charge fluctuations in multilevel quantum dots in and out of equilibrium}

\author{Carsten J. Lindner}
\affiliation{Institut f\"ur Theorie der Statistischen Physik, RWTH Aachen, 
52056 Aachen, Germany and JARA - Fundamentals of Future Information Technology}
\author{Fabian B. Kugler}
\affiliation{Physics Department, Arnold Sommerfeld Center for Theoretical Physics, and Center
for NanoScience,Ludwigs-Maximilians-Universit\"at M\"unchen, Theresienstr. 37, 80333 Munich, Germany}
\author{Volker Meden}
\affiliation{Institut f\"ur Theorie der Statistischen Physik, RWTH Aachen, 
52056 Aachen, Germany and JARA - Fundamentals of Future Information Technology}
\author{Herbert Schoeller}
\email[Email: ]{schoeller@physik.rwth-aachen.de}
\affiliation{Institut f\"ur Theorie der Statistischen Physik, RWTH Aachen, 
52056 Aachen, Germany and JARA - Fundamentals of Future Information Technology}

\begin{abstract}
{We present the real-time renormalization group (RTRG) method as a method to describe the stationary state current through generic multi-level quantum dots in nonequilibrium. The employed approach consists of a very rudiment approximation for the RG equations which neglects all vertex corrections while it provides a means to compute the effective dot Liouvillian self-consistently. Being based on a weak-coupling expansion in the tunneling between dot and reservoirs, the RTRG approach turns out to reliably describe charge fluctuations in and out of equilibrium for arbitrary coupling strength, even at zero temperature. We confirm this in the linear response regime with a benchmark against highly-accurate numerically renormalization group data in the exemplary case of three-level quantum dots. For small to intermediate bias voltages and weak Coulomb interactions, we find an excellent agreement between RTRG and functional renormalization group data, which can be expected to be accurate in this regime. As a consequence, we advertise the presented RTRG approach as an efficient and versatile tool to describe charge fluctuations theoretically in quantum dot systems.}

\end{abstract}


\maketitle

\section{Introduction}
\label{sec:introduction}
Describing electron transport through mesoscopic systems like semiconductor heterostructures\cite{review_qi} or molecules (e.g. carbon nanotubes\cite{review_nanotubes}) at low temperatures in nonequilibrium is a fundamental problem in the field of quantum statistics. The physics of these systems is highly affected by the repulsive Coulomb interaction between the electrons, leading to interesting correlation phenomena like the Kondo effect\cite{kondo_theo,kondo_exp}. Further attraction arose from possible applications of quantum nanostructures in future information technology, in particular in quantum computers.

Two competing mechanisms drive the physical behavior of an open quantum dot. First, electrons can tunnel in and out  of the quantum dot via tunnel barriers, separating the dot from surrounding reservoirs held at different temperatures and chemical potentials. Second, the occupancy of the dot by the electrons is highly affected by the Pauli principle in concert with the repulsive Coulomb interaction between the electrons. The interplay of these two mechanisms causes correlation effects resulting in emergent phenomena such as the Kondo effect at sufficiently low temperatures. 

Transport spectroscopy provides a means to analyze the physical processes in open quantum dots\cite{review_qi,andergassen_etal}. The idea is to scrutinize the current through the quantum dot as function of the bias voltage, gate voltage or external magnetic fields. For instance, a resonance peak in the linear conductance as function of the gate voltage signals the change of the average dot electron number\cite{review_qi,andergassen_etal}, while the emergence of a plateau is a hallmark of the Kondo effect\cite{glazman_pustilnik_05}. In contrast, an increase in the step-like current away from equilibrium indicates the opening up of another transport channel, i.e. the possibility of occupying an excited state of the quantum dot\cite{review_qi,andergassen_etal}. Finding adequate approaches and methods to theoretically describe resonances in the current through nanostructures is therefore of great interest.

In equilibrium, numerically exact methods such as the numerical renormalization group (NRG)\cite{Wilson1975,Bulla2008} or the density matrix renormalization group (DMRG)\cite{review_dmrg} are well-established to describe the current through quantum nanostructures. Some progress has also been made in order to generalize these approaches to nonequilibrium, leading to the scattering state NRG\cite{ST-NRG}, time-dependent NRG (TD-NRG)\cite{TD-NRG} and the time-dependent DMRG (TD-DMRG)\cite{TD-DMRG}. Recently, a novel thermofield approach\cite{Schwarz2018} was developed that combines the latter two methods to describe impurity models in nonequilibrium. Although all these approaches are very promising, reliable numerical data for the current across generic quantum dots with more than two levels out of equilibrium is missing in the literature at the moment. 

Numerically exact methods are typically computationally demanding and one therefore often assumes certain symmetries for the model to reduce the numerical effort. Essentially analytic methods such as the real-time RG (RTRG)\cite{schoeller_2009,schoeller_2014}, functional RG (fRG)\cite{frg_review,jakobs_meden_schoeller_2007,jakobs_pletyukhov_schoeller_2010a} or the flow equation method\cite{flow_equation_review} are usually less demanding, allowing for a more efficient study of complex setups. For instance, the computational effort for determining the self-energy using the fRG method in lowest order truncation is comparable to that of a mean-field calculation.

The downside of analytic methods is that they are usually perturbative with the consequence that their range of applicability is restricted. However, perturbative RG methods as the fRG or the RTRG are based on a resummation of certain classes of diagrams. If these diagrams capture the essential physical processes, then these methods yield reliable results even beyond the range of validity of a corresponding approximation within plain perturbation theory. A notable example in this regard is the agreemenent between results for the Kondo model in nonequilibrium in the strong-coupling limit obtained from a RTRG approach\cite{pletyukhov_schoeller_2012,reininghaus_pletyukhov_schoeller_2014}, which is perturbative in the coupling between dot and reservoirs, and exact numerical methods \cite{Schwarz2018}. Some results are also in accordance with experimental data\cite{strong_coupling_exp}.

In this article, we report a similar observation for the description of charge fluctuations in generic three-level quantum dots with nondegenerate single-particle energies. 
Hereby, the regime of charge fluctuations is defined by the condition that real processes are possible changing the particle number on the quantum dot by $\Delta N=\pm 1$. In this regime, Kondo induced correlations (as discussed in Ref.~\onlinecite{effective_tunneling_models} for the Coulomb blockade regime) are suppressed and the main physics consists in resonances for the differential conductance as function of the gate voltage when one of the renormalized single-particle excitations of the dot is close to one of the chemical potentials of the reservoirs. Such resonances occur also in the sequential tunneling regime of high temperatures $T\gg\Gamma$, where $\Gamma$ denotes the broadening of the single-particle excitations induced by the coupling to the leads. In this regime, the resonance positions correspond to the bare single-particle excitations of the dot and their line shape is mostly dominated by thermal smearing. This can be described by standard kinetic equations in Born-Markov approximation. In contrast, the aim of the present paper is to calculate the position and line shape of these resonances at zero temperature $T=0$ by including all diagrams of the RTRG describing charge fluctuation processes. In this essentially non-perturbative regime in $\Gamma$ one obtains renormalized resonance positions and the line shape is dominated by quantum fluctuations leading to Breit-Wigner like line shapes with a broadening of the order of $\Gamma$. Since orbital fluctuations are not taken into account the solution is expected to be reliable when the distance $\delta$ of the gate voltage to one of the resonance positions is of the order of $\Gamma$. Furthermore, since the RTRG is derived from a diagrammatic expansion in $\Gamma$, at first glance this method is controlled only for small dot reservoir couplings, which means that $\Gamma$ should be smaller than $\text{max}\{T,\delta\}$. However, our study reveals that the self-consistent resummation of all charge fluctuation diagrams via the RTRG approach yields reliable results close to the resonances for arbitrary Coulomb interactions and arbitrary coupling to the reservoirs, respectively, even at zero temperature. Even when all energy scales become of the same order of magnitude $\delta,U\sim\Gamma$, where one can no longer distinguish between the regime of charge fluctuations (close to the resonances) and orbital fluctuations (between the resonances), the considered RTRG approximation describes quite well the line shape of the main resonances but not the conductance between the resonances (where orbital fluctuations dominate).
This means a drastic extension of the range of validity of this approximation. To confirm this, highly-accurate NRG data for the linear conductance as function of the gate voltage serves as a benchmark against the RTRG solution. In nonequilibrium, we find an excellent agreement between the fRG method, which employs the Coulomb interaction as the expansion parameter, and the RTRG for small Coulomb interactions and strong coupling, respectively. Additionally, one can show that our approximate RTRG approach becomes exact for large bias voltages, see Appendix \ref{app:perturbation_theory}. As a consequence, we advertise the RTRG method as an efficient tool to describe charge fluctuations in multi-level quantum dots in nonequilibrium even at very low temperatures.

The fRG in static approximation serves in the following mainly as a benchmark for small Coulomb interactions in nonequilibrium, where this approach is strictly controlled. However, previous studies of transport through multi-level quantum dots with a complex setup\cite{frg-eq} revealed that the fRG is reliable up to intermediate Coulomb interactions in the linear response regime. In general, fRG in static approximation is applicable if the physical behavior can be described by an effective single-particle picture. While this is clearly not the case for large bias, we compare fRG and RTRG data for the differential conductance also in this regime in order to estimate the range of applicability for the effective single-particle picture.

In this article, we stick to simple approximation schemes for the RTRG and the fRG in order to keep the numerical effort as low as possible. However, both methods are flexible in the sense that approximations can be systematically extended by taking higher order diagrams into account, as it was demonstrated, e.g., for a theoretical description of two-level quantum dots by the RTRG\cite{saptsov_wegewijs_2012} and the fRG\cite{jakobs_pletyukhov_schoeller_2010a,karrasch_et_al_2008}.

The outline of this article is as follows. In section \ref{subsec:anderson_model}, we introduce the multi-level generalization of the Anderson model together with a generic model for the tunneling between dot and reservoirs. The considered methods - RTRG, fRG and NRG - are then introduced successively in sections \ref{subsec:rtrg}-\ref{subsec:nrg}. Section \ref{sec:linear_response} comprise the benchmark of the considered RTRG and fRG approximations against NRG data for the linear conductance for a model with proportional coupling. Afterwards, we discuss the reliability of the RTRG and fRG approaches to describe the quantum dot with generic tunneling matrix in nonequilibrium in section \ref{sec:nonequilibrium}. The article closes with a summary of the main results. We consider $\hbar\,=\,k_\mathrm{B}\,=\,e\,=\,1$ throughout this article.

\section{Model and Methods}
\label{sec:model}

In this section, we briefly introduce the considered model for the quantum dot as well as the methods applied in this work. To this end, we first discuss the Anderson model for multi-level quantum dots. Then, we set up the RG equations for this model using the RTRG method with the reservoir-dot couplings being the expansion parameter. Similarly, we set up RG equations in the static approximation within the fRG approach with the Coulomb interaction being the expansion parameter and comment on the applied NRG method. Results from the fRG are later on used to test the reliability of the RTRG solution out of equilibrium in the regime of weak Coulomb interactions and strong coupling,while the highly-accurate NRG data provides a benchmark for the linear conductance at arbitrary Coulomb interactions.

\subsection{Multi-level Anderson model}
\label{subsec:anderson_model}

We consider the multi-level generalization\cite{bickers_1987} of the single impurity Anderson model\cite{anderson_1961} where the electron spin index $\sigma$ is replaced by the flavor index $l$. This is a quantum number labeling one of the $Z$ dot levels which is either empty or is occupied by exactly one electron. In general, $l$ can be viewed as a multi-index that also includes the spin index $\sigma$. The corresponding Hamiltonian reads as
\begin{align}
\label{eq:H_s}
H_\mathrm{s} \,&=\, H_0 \, + \, V_\mathrm{ee} \quad, \\
\label{eq:H_0}
H_0 \,&=\,  \sum_{l}\, \varepsilon_{l}\, c_{l}^{\dagger}\, c_{l} \quad, \\
\label{eq:V_ee}
V_\mathrm{ee} \,&=\, \frac{U}{2}\, \sum_{ll^\prime}\, c_{l}^{\dagger}\, c_{l^\prime}^{\dagger}\, c_{l^\prime}\,  c_{l} \quad,
\end{align}
Here, $U$ quantifies the strength of the Coulomb interaction between the dot electrons and $\varepsilon_l=h_l - V_\mathrm{g} - (Z-1) {U \over 2}$ are the single-particle dot levels. To avoid a proliferation of parameters we assume a flavor independent Coulomb interaction. However, our approaches can also handle more general two-particle interaction terms by incorporating these terms into the initial conditions of the RG equations. External fields (e.g. magnetic fields) are incorporated into the level spacing $h_l$ and $V_\mathrm{g}$ is the gate voltage, allowing to uniformly tune the dot levels. The choice $h_l=V_\mathrm{g}=0$ defines the particle-hole symmetric model. 

The full Hamiltonian of the $Z$-level Anderson model is given by
\begin{align}
\label{eq:H_tot}
H_{\mathrm{tot}} \,&=\, H_{\mathrm{s}}\, +\, H_ {\mathrm{res}}\, + V_\mathrm{c}\quad,
\end{align}
with
\begin{align}
\label{eq:H_res}
H_{\mathrm{res}} \,&=\,\sum_{k \alpha l} \, \varepsilon_{k \alpha l}\, a_{k \alpha l}^{\dagger}\, a_{k \alpha l} \quad,\\
\label{eq:V_c}
V_\mathrm{c} \,&=\, {1\over\sqrt{\rho^{(0)}}}\,\sum_{k \alpha l l^\prime}\, \left(t^{\alpha}_{l l^\prime} a^\dagger_{k \alpha l} c_{l^\prime} \,+\, (t^{\alpha}_{l l^\prime})^* c^\dagger_{l^\prime} a_{k \alpha l} \right)\quad,
\end{align}
where $H_{\mathrm{res}}$ is the part accounting for the $Z_\mathrm{res}$ reservoirs and $V_\mathrm{c}$ the coupling between the quantum dot and the reservoirs. Accordingly, $\alpha=1,\ldots\,Z_\mathrm{res}$ is the reservoir index, $\varepsilon_{k \alpha l}$ the band dispersion relative to the chemical potential $\mu_\alpha$ for the channel $l$ with some quantum number $k$ that becomes continuous in the thermodynamic limit. Furthermore, $t^{\alpha}_{l l^\prime}$ denotes the matrix elements of the tunneling between the reservoir and the dot. We assume flat reservoir bands (at least on the low-energy scale of interest) and take $t^{\alpha}_{l l^\prime}$ as independent of $k$. Here, $\rho^{(0)}$ is some average reservoir density of states which we set to $\rho^{(0)}=1$ for convenience, defining the energy units.

The reservoirs contribute to the self-energy and the current formula only via the \textit{hybridization matrix}
\begin{align}
\label{eq:hybridization_matrix}
\Gamma^\alpha_{ll^\prime}(\omega) \,=\, 2\pi\, \sum_{l_l l_2}\, (t^{\alpha}_{l_1 l})^*\, \rho^\alpha_{l_1 l_2} (\omega) \,t^{\alpha}_{l_2 l^\prime}\quad,
\end{align}
where $\rho^\alpha_{l_1 l_2} (\omega)\,=\, \delta_{l_1 l_2} \sum_k \delta(\omega - \varepsilon_{k \alpha l_1} + \mu_\alpha)$ is the constant density of states in reservoir $\alpha$. This together with the assumption that the reservoirs are infinitely large means that we can neglect the frequency dependence of $\Gamma^\alpha_{ll^\prime}(\omega)$. In particular, we consider the \textit{normal lead model} with
\begin{align}
\label{eq:normal_lead_model}
\Gamma^\alpha_{ll^\prime} \,=\, 2\pi \,\sum_{l_l}\, (t^{\alpha}_{l_1 l})^* \,t^{\alpha}_{l_1 l^\prime}\quad,
\end{align}
in the following. We define $\Gamma=\sum_{\alpha ll^\prime} \Gamma^\alpha_{ll^\prime}$ as the characteristic energy scale for tunneling processes between the dot and the reservoirs.

Importantly, the dot expectation values and the current depend on the form of the hybridization matrices and \textit{not} on the form of the tunneling matrices. This means that different models with the same hybridization matrices have the same properties. Accordingly, all these models can be mapped onto each other with rotations in the channel indices with an invariant hybridization matrix \cite{effective_tunneling_models}. This is the reason why we can describe the generic case using the normal lead model \eqref{eq:normal_lead_model} where the dot and channel indices coincide.

Finally, the Fermi distribution
\begin{align}
\label{eq:Fermi_distribution}
f_\alpha (\omega) \,=\, \frac{1}{e^{\beta_\alpha \omega}+1}
\end{align}
characterizes the thermodynamic state of the reservoir with the inverse temperature $\beta_\alpha=T_\alpha^{-1}$. We later consider resevoir temperatures $T_\alpha=0$ implying $f_\alpha (\omega)\,=\,\Theta(\omega)$ for the Fermi distribution function with $\Theta(\omega)$ being the Heaviside distribution.

\subsection{Real-time RG}
\label{subsec:rtrg}

The state of the quantum dot can be quantified by the reduced density matrix
\begin{align}
\label{eq:reduced_density_matrix}
\rho_\mathrm{s}(t) \,=\, \tr_\mathrm{res} \rho_\mathrm{tot}(t),
\end{align}
where $\tr_\mathrm{res}$ is the trace over the reservoir degrees of freedom and the total density matrix $\rho_\mathrm{tot}(t)$ is the solution of the von Neumann equation $i \td{}{t} \rho_\mathrm{tot}(t)\,=\, \com{H_\mathrm{tot}}{\rho_\mathrm{tot}(t)}$. The reduced density matrix $\rho_\mathrm{s}(t)$ is in turn the solution of the kinetic equation
\begin{align}
\label{eq:kinetic_equation}
i \td{}{t} \rho_\mathrm{s}(t) \,=\, \int\limits_{0}^t \mathrm{d}t^\prime\,L(t-t^\prime) \rho_\mathrm{s}(t^\prime) 
\end{align}
with the effective Liouvillian $L(t-t^\prime)$ being the response function due to the coupling to the reservoirs. This equation can be formally solved in Fourier space, yielding
\begin{align}
\label{eq:formal_solution_Fourier_space}
\rho_\mathrm{s}(E) \,=\, \frac{i}{E - L(E)} \rho_\mathrm{s}(t=0) 
\end{align} 
with $\rho_\mathrm{s}(E)\,=\, \int_{0}^\infty \mathrm{d}t\,e^{iEt}\,\rho_\mathrm{s}(t)$ and $L(E)\,=\, \int_{0}^\infty \mathrm{d}t\,e^{iEt}\,L(t)$. 

Here, we are only interested in the solution in the stationary limit ($t \rightarrow \infty$) which is defined as $\rho_\mathrm{st} = \lim_{E \rightarrow i0^+} (-iE) \rho_\mathrm{s}(E)$. It can be conveniently obtained from solving
\begin{align}
\label{eq:stationary_state}
L(i0^+) \rho_\mathrm{st} \,=\, 0 \quad. 
\end{align} 

The average electron current leaving reservoir $\gamma$ is defined as $I_\gamma(t)\,=\, \expect{- \td{}{t} \hat{N}_\gamma}$, where $\hat{N}_\gamma=\sum_{k l} a^\dagger_{k \gamma l} a_{k \gamma l}$ is the particle number in reservoir $\gamma$. The current can conveniently be computed using
\begin{align}
\label{eq:current_rtrg}
I_\gamma(t)\,=\ -i\, \int\limits_{0}^{t} \mathrm{d}t^\prime \,\tr_\mathrm{s} \Sigma_\gamma(t-t^\prime)\rho_\mathrm{s}(t^\prime) \quad,
\end{align}
or in Fourier space
\begin{align}
\label{eq:current_rtrg_Fourier}
I_\gamma(E)\,=\ -i\, \tr_\mathrm{s} \Sigma_\gamma(E)\,\rho_\mathrm{s}(E) \quad,
\end{align}
where $\Sigma_\gamma(t-t^\prime)$ and $\Sigma_\gamma(E) \,=\, \int_{0}^{\infty} \mathrm{d}t \, e^{iEt} \Sigma_\gamma(t)$, respectively, is the current kernel. The stationary state limit is given by
\begin{align}
\notag
I_\gamma^\mathrm{st} \,&=\, \lim_{E \rightarrow i0^+} (-iE) I_\gamma(E) \\
\label{eq:current_rtrg_stationary_state}
&=\, -i\, \tr_\mathrm{s} \Sigma_\gamma(i0^+)\,\rho_\mathrm{st} \quad,
\end{align}
which we aim to compute.

The model Hamiltonian (\ref{eq:H_0}-\ref{eq:V_c}) provides two different starting points for a perturbative expansion. First, for weak Coulomb interactions ($U \ll \Gamma$), $V_\mathrm{ee}$ can be viewed as a perturbation and one  can expand in the electron-electron interaction. This is the starting point of the fRG that is discussed in section \ref{subsec:frg}. Second, for arbitrary $U$, a weak-coupling expansion in $\Gamma$ is favorable for $\Gamma\ll\max\{T_\alpha,\delta\}$. In this case, one can compute the effective Liouvillian $L(E)$ and the current kernel $\Sigma_\gamma(E)$ using the RTRG approach, as we discuss now.

Applying the diagrammatic technique presented in Refs.~\onlinecite{schoeller_2009,schoeller_2014} on Anderson-type models with charge fluctuations yields the RG equation
\begin{align}
\notag
\td{}{E} L(E) \,&=\, - \, \begin{tikzpicture}[baseline=0.5ex,scale=0.5]
\node[node1] (left) at (-1.25,0) {};
\node[node1] (right) at (1.25,0) {};
\draw[line1,black] (left) -- (right) {};
\draw[line1,ForestGreen] (left) -- (-1.25,1) -- (1.25,1) -- (right) {};
\node[node4] (derrivative) at (0,1)  {};
\end{tikzpicture} \,+\, \orderof{G^4} \quad,\\
\notag
&=\, - \int \mathrm{d}\omega\, f^\prime(\omega)\, G_1(E,\omega)\\
\label{eq:liouvillian_rg_starting_point}
&\quad \times \,\Pi(E_1+\omega)\,G_{\overline{1}}(E_1+\omega,-\omega) + \orderof{G^4}
\end{align}
for the effective Liouvillian, which was also already stated in the supplementary material of Ref.~\onlinecite{kashuba_schoeller_2013}. Here, $\Pi(E)\,=\,i[E-L(E)]^{-1}$ is the full propagator of the quantum dot and $G_1(E,\omega)$ is an effective vertex, accounting for the dot-reservoir interaction. Furthermore, $E_1=E+\overline{\mu}_1$ is the Fourier variable plus the chemical potential $\overline{\mu}_1=\eta \mu_\alpha$, $1=\eta \alpha l$ is a multi-index and $\eta$ is a sign index that indicates whether a dot electron is created or annihilated during the interaction process. Accordingly, $\eta=+$ ($\eta=-$) corresponds to the dot annihilation (creation) operator, i.e. $c_{+l}=c_l$ ($c_{-l}=c_l^\dagger$).

The derivation of the RG equation (\ref{eq:liouvillian_rg_starting_point}) is not very difficult and can be sketched as follows, for details see Refs.~\onlinecite{pletyukhov_schoeller_2012,kashuba_schoeller_2013,schoeller_2014}. First, the perturbative series for $L(E)$ consists of a series of bare vertices $G_1$ connected by bare propagators $\Pi^{(0)}(E+X)=i[E+X-L_0]^{-1}$, where $L_0=[H_0,\cdot]$ is the Liouvillian of the bare dot and $X$ contains a certain sum of chemical potentials and frequencies of the reservoir contractions connecting the bare vertices. After resummation of self-energy insertions all bare propagators are replaced by the full effective ones $\Pi(E+X)$. Differentiating this series w.r.t. $E$ means that one of the propagators is replaced by its derivative $\td{}{E}\Pi(E+X)$. Resumming vertex corrections left and right to $\td{}{E}\Pi(E+X)$ and considering only the charge fluctuation process yields to lowest order 
\begin{align}
\notag
\td{}{E} L(E) 
&=\,  \int \mathrm{d}\omega\, f(\omega)\, G_1(E,\omega)\\
\label{eq:liouvillian_rg_derivation}
&\hspace{-1cm}
\times \,\td{}{E}\,\Pi(E_1+\omega)\,G_{\overline{1}}(E_1+\omega,-\omega)\,+\,O(G^4)\,.
\end{align}
Using $\td{}{E}\,\Pi(E_1+\omega)=\td{}{\omega}\,\Pi(E_1+\omega)$ and partial integration one can shift the frequency derivative to the Fermi function and to the effective vertices. Since one can show that the frequency derivative of the vertices again leads to higher order terms, they can be neglected and one obtains the RG equation (\ref{eq:liouvillian_rg_starting_point}).

The effective vertex $G_1(E,\omega)$ can be obtained as the solution of a similar RG equation. However, as it is explained in appendix \ref{app:perturbation_theory}, a resummation of logarithmic terms in the perturbative series expansion is not necessary since the self-consistently calculated Liouvillian does not suffer from any logarithmically divergent terms for $E=i0^+$. This has the consequence that vertex corrections can be neglected in leading order and we can replace the effective vertices $G_1(E,\omega)$ by the bare ones, i.e.
\begin{align}
\label{eq:vertex_symmetric}
G_1 \,&=\, \sum_p G_1^p
\end{align}
with
\begin{align}
\label{eq:vertex_liouvillian}
G^p_1\,=\,G^p_{\eta \alpha l}\,&=\, \sum_{l^\prime} t^{\eta \alpha}_{l l^\prime} C^p_{\eta l^\prime} \quad,
\end{align}
where $t^{\eta \alpha}_{l l^\prime} \,=\,\delta_{\eta +}\,t^{\alpha}_{l l^\prime}\,+\,\delta_{\eta -}\,t^{\alpha}_{l^\prime l}\,=\,t^{-\eta \alpha}_{l^\prime l}$ and
\begin{align}
\label{eq:dot_field_superoperator}
C_{\eta l}^{p} \,\bullet \,&=\, p \sigma^{p} \begin{cases}
c_{\eta l}\, \bullet & \text{if} \quad p=+ \\
\bullet\, c_{\eta l} & \text{if} \quad p=-
\end{cases}
\end{align}
are the dot field superoperators fulfilling the anticommutation relation $\anticom{C_{\eta l}^{p}}{C_{\eta^\prime l^\prime}^{p^\prime} } \,=\, p \delta_{p p^\prime} \delta_{\eta, - \eta^\prime} \delta_{l l^\prime}$. Here, the sign factor $\lbra{s_1 s_2} \sigma^p \lket{s_1^\prime s_2^\prime} \,=\, \delta_{s_1s_1^\prime} \delta_{s_2s_2^\prime} p^{N_{s_1} - N_{s_2}}$ measures the parity of the states\cite{schoeller_2009,schoeller_2014} $\lket{s s^\prime}= \ket{s}\bra{s^\prime}$, where $\lket{s s^\prime}= \ket{s}\bra{s^\prime}$ are the basis states of the dot Liouville space, $\lbra{ss^\prime}\ldots=\bra{s} \ldots \ket{s^\prime}$ are the basis states of the corresponding dual Liouville space, $\ket{s}$ are the many-body eigenstates of $H_\text{s}$ and $N_s$ the dot electron number in state $\ket{s}$.

A similar RG equation for the current kernel follows from \eqref{eq:liouvillian_rg_starting_point} by simply replacing the left vertex $G_1(E,\omega)$ by the current vertex $(I_\gamma)_1(E,\omega)$. This yields
\begin{align}
\notag
\td{}{E} \Sigma_\gamma(E) \,&=\, - \int \mathrm{d}\omega\, f^\prime(\omega)\, (I_\gamma)_1(E,\omega)\\
\label{eq:current_kernel_rg_starting_point}
&\quad \times \,\Pi(E_1+\omega)\,G_{\overline{1}}(E_1+\omega,-\omega) \quad.
\end{align}
For the same reasons as above, we neglect the vertex corrections to the current kernel which means that we insert\cite{schoeller_2009,schoeller_2014}
\begin{align}
\label{eq:current_vertex}
(I_\gamma)_1,&=\, \sum_{p=\pm} (I_\gamma)^p_1 \,=\, c^\gamma_1 \tilde{G}_1
\end{align}
for the current vertex, where $c^\gamma_1  \,=\, c^\gamma_{\eta \alpha}  \,=\, - \frac{1}{2} \eta \delta_{\alpha \gamma}$ and $\tilde{G}_1 \,=\, \sum_{p=\pm} p G^p_1$.

The RG flow starts at $E=iD$, with $D$ being the band-width of the reservoir density of states (see Appendix \ref{app:perturbation_theory}), and stops at $E=i0^+$, where the effective Liouvillian and the current kernel needed to compute the stationary state properties are defined. Setting up the initial conditions for the RG equations as explained in Ref.~\onlinecite{reininghaus_pletyukhov_schoeller_2014}, we obtain
\begin{align}
\label{eq:liouvillian_initial}
L(E) \big|_{E=iD} \,&=\, L^{(0)} + L^{(1\mathrm{s})} \quad,\\
\label{eq:current_kernel_initial}
\Sigma_\gamma(E) \big|_{E=iD} \,&=\, \Sigma^{(1\mathrm{s})}_{\gamma} \quad,
\end{align}
from lowest-order perturbation theory where $L^{(1\mathrm{s})}$ and $\Sigma^{(1\mathrm{s})}_{\gamma}$ are given by \eqref{eq:liouvillian_symmetric_part} and \eqref{eq:current_kernel_symmetric_part}. The natural choice for the path of the RG flow is $E=i \Lambda$ with $D \geq \Lambda \geq 0^+$ and a {\it real} flow parameter. This is a special choice since, in general, the flow parameter $E$ within the $E$-flow scheme\cite{pletyukhov_schoeller_2012,reininghaus_pletyukhov_schoeller_2014} of the RTRG is complex with the consequence that two different paths for the RG flow connecting the same starting and end point yield the same solution at the end point, as long as they do not enclose any singularities of $L(E)$ and $\Sigma_\gamma(E)$, which lie in the lower half of the complex plane. This is fundamental for computing the transient dynamics\cite{schoeller_2014,kashuba_schoeller_2013}.

At zero temperature, $T_\alpha=0$, the derivative of the Fermi distribution becomes the $\delta$-distribution, $f^\prime(\omega)\,=\,-\delta(\omega)$, and the frequency integrals in \eqref{eq:liouvillian_rg_starting_point} and \eqref{eq:current_kernel_rg_starting_point} become trivial. Thus, we obtain
\begin{align}
\notag
&\td{}{\Lambda} \tilde{L}(\Lambda) \,= \\
\label{eq:liouvillian_rg_real_flow_parameter}
&=\, i \sum_{\eta \alpha l} G_{\eta \alpha l}\, \frac{1}{i \Lambda + \overline{\mu}_\alpha - \tilde{L}(\Lambda - i\, \overline{\mu}_\alpha)} \, G_{-\eta \alpha l} \quad, \\
\notag
&\td{}{\Lambda} \tilde{\Sigma}_\alpha (\Lambda) \,= \\
\label{eq:current_kernel_rg_real_flow_parameter}
&=\, - \frac{i}{2} \sum_{l \eta} \eta\, \tilde{G}_{\eta \alpha l} \frac{1}{i \Lambda + \overline{\mu}_\alpha - \tilde{L}(\Lambda - \, i\overline{\mu}_\alpha)} G_{-\eta \alpha l} \quad,
\end{align}
with $\tilde{L} (\Lambda) \,=\, L(i\Lambda)$ and $\tilde{\Sigma} (\Lambda) \,=\, \Sigma(i\Lambda)$.

We note that \eqref{eq:liouvillian_rg_real_flow_parameter} defines an infinite hierarchy of differential equations since the Liouvillian evaluated at $\Lambda - i \overline{\mu}_\alpha$ is fed back and \textit{not} the one evaluated at $\Lambda$. Thus, one also needs to solve an RG equation for $\tilde{L}(\Lambda - \, i\overline{\mu}_\alpha)$. The right hand side of this equation in turn depends on $\tilde{L}(\Lambda - i \overline{\mu}_\alpha - i \overline{\mu}_{\alpha^\prime})$. By proceeding this way, we arrive at an infinite hierarchy of RG equations for the effective Liouvillian where each RG equation is associated with a different shift in the energy argument of the effective Liouvillian. However, this hierarchy of RG equations can be straightforwardly truncated, as explained in Appendix \ref{app:truncation}.

In total, the purpose of the RG treatment is a {\it self-consistent} computation of the effective Liouvillian $\tilde{L}(\Lambda)$. This is necessary since bare perturbation theory for the Liouvillian and the current kernel exhibits logarithmic singularities, see the discussion in Appendix \ref{app:perturbation_theory}. These singularities are located at
\begin{align}
\label{eq:resonant_tunneling_pt}
\mu_\alpha \,=\, E_{s_1} - E_{s_2} \quad \text{with}\, N_{s_1} = N_{s_2} + 1 \quad,
\end{align}
where $E_{s}$ are the eigenvalues of $H_\mathrm{s}$. This equation represents the well-known condition for resonant tunneling through the quantum dot, see, e.g., Refs.~\onlinecite{review_qi,andergassen_etal} for a review. This means that the logarithmic singularities result in $\delta$-peaks in the differential conductance, i.e. the derivative of the current with respect to the reservoir bias voltage. As a consequence of the RG treatment, the eigenvalues $\lambda_k(E)$ of the effective Liouvillian, defined by $L(E) \lket{x_k(E)} \,=\, \lambda_k(E) \lket{x_k(E)}$, replace $E_{s_1} - E_{s_2}$ in the argument of the logarithms of $L(E)$ and $\Sigma_\gamma(E)$. Importantly, the imaginary part $\imag \lambda_k(i0^+)$ provides a cut-off in the argument of the logarithm. This regularizes the logarithmic singularities and causes a finite height of the conductance peaks together with a finite broadening of width $\sim \Gamma$. In addition, the peak position is renormalized, i.e. the conductance peaks are now located at
\begin{align}
\label{eq:resonant_tunneling_rtrg}
\overline{\mu}_\alpha - \real \lambda_k(\overline{\mu}_\alpha + i 0^+) \,=\, 0 \quad.
\end{align}

This must be contrasted to the case of moderate temperatures $T_\alpha \gg \Gamma$, where, e.g., the width of the conductance peaks is given by the temperature $T\,=\,T_\alpha$ if all reservoir temperatures are equal. In this case, the sharp edge of the Fermi distribution, being fundamental for the emergence of logarithmic singularities at $T_\alpha=0$, is broadened by the temperature and no logarithmic singularities occur. In this case, the full propagator on the right hand sides of the RG equations \eqref{eq:liouvillian_rg_starting_point} and \eqref{eq:current_kernel_rg_starting_point} can be replaced by the bare one $\Pi^{(0)}(E)\,=\,i[E- L^{(0)}]^{-1}$, where $L^{(0)}$ is given by \eqref{eq:liouvillian_zeroth_order}. Thus, the RG equations can be formally solved, yielding the expressions for the first-order corrections in bare perturbation theory.

\subsection{Functional RG}
\label{subsec:frg}
An alternative approach to compute the current across the quantum dot is the Keldysh Green's function formalism\cite{keldysh_1965}. The current can be computed from
\begin{align}
\notag
I^\text{st}_\gamma \,&=\, \frac{i}{4 \pi} \int \mathrm{d} \omega \, \tr \dunderline{\Gamma}^\gamma \left\lbrace \left[1 - 2 f_\gamma (\omega - \mu_\gamma) \right] \right. \\
\label{eq:current_frg}
&\quad \left. \times \left[\dunderline{G}^\text{R} (\omega) - \dunderline{G}^\text{A} (\omega) \right] - \dunderline{G}^\text{K} (\omega) \right\rbrace \quad,
\end{align}
which is straightforwardly obtained from the current formula stated in Ref.~\onlinecite{meir_wingreen_92} by replacing the lesser component of the Green's function by the Keldysh component. Accordingly, $\dunderline{G}^{\mathrm{R},\mathrm{A},\mathrm{K}} (\omega)$ is the retarded, advanced and Keldysh component of the dot Green's function, respectively,
\begin{align}
\label{eq:greens_function_matrix}
\dunderline{G}(\omega) \,=\, \left(\begin{array}{cc}
\dunderline{G}^{\mathrm{R}} (\omega) & \dunderline{G}^{\mathrm{K}} (\omega) \\
0 & \dunderline{G}^{\mathrm{A}} (\omega)
\end{array} \right)  \quad,
\end{align}
and $\dunderline{\Gamma}^\gamma$ is the hybridization matrix in matrix notation, i.e. $(\dunderline{\Gamma}^\gamma)_{ll^\prime}\,=\,\Gamma^\gamma_{ll^\prime}$. There are in total two independent components of the Green's function, that are
\begin{align}
\label{eq:retarded_greens_function}
\dunderline{G}^{\mathrm{R}} (\omega) \,&=\, \frac{1}{\omega - \dunderline{\Sigma}^{\mathrm{R}} (\omega)} \,=\, [\dunderline{G}^{\mathrm{A}}(\omega)]^\dagger \quad, \\
\label{eq:keldysh_greens_function}
\dunderline{G}^{\mathrm{K}} (\omega) \,&=\, \dunderline{G}^{\mathrm{R}} (\omega) \dunderline{\Sigma}^{\mathrm{K}}(\omega) \dunderline{G}^{\mathrm{A}}(\omega) \quad,
\end{align}
where
\begin{align}
\label{eq:selfenergy_matrix}
\dunderline{\Sigma}(\omega) \,=\, \left(\begin{array}{cc}
\dunderline{\Sigma}^{\mathrm{R}} (\omega) & \dunderline{\Sigma}^{\mathrm{K}} (\omega) \\
0 & \dunderline{\Sigma}^{\mathrm{A}} (\omega)
\end{array} \right)
\end{align}
is the self-energy.

Here, we already consider the so-called {\it reservoir dressed} Green's function. This is an effective Green's function in dot space, hence doubly underlined in the matrix notation, which can be obtained from the Green's function of the total system by projecting out the reservoir degrees of freedom. The projection results in an additional addend to the self-energy in terms of $\dunderline{\Gamma}^\alpha$. In the non-interacting case, i.e., $U=0$, we obtain $\dunderline{\Sigma}^{\mathrm{R}}\,=\,\dunderline{\varepsilon}+\dunderline{\Sigma}^{\mathrm{R}}_{\mathrm{res}}\,=\,(\dunderline{\Sigma}^{\mathrm{A}})^\dagger$ and $\dunderline{\Sigma}^{\mathrm{K}}(\omega)\,=\,\dunderline{\Sigma}^{\mathrm{K}}_{\mathrm{res}}(\omega)$ with $(\dunderline{\varepsilon})_{ll^\prime}\,=\,\varepsilon_{ll^\prime}$ and
\begin{align}
\label{eq:retarded_reservoir_selfenergy}
\dunderline{\Sigma}^\mathrm{R}_\mathrm{res} \,&=\, -\frac{i}{2} \dunderline{\Gamma} \quad,\\
\label{eq:keldysh_reservoir_selfenergy}
\dunderline{\Sigma}^\mathrm{K}_\mathrm{res}(\omega) \,&=\, -i \sum_\alpha \bigg[1 - 2 f_\alpha(\omega-\mu_\alpha)\bigg] \dunderline{\Gamma}^\alpha \quad,
\end{align}
with $\dunderline{\Gamma}\,=\,\sum_\alpha \dunderline{\Gamma}^\alpha$. Accordingly, the reservoir dressed Green's function of the non-interacting system $(U=0)$ is given by
\begin{align}
\label{eq:retarded_greens_function_noninteracting}
\dunderline{G}^{\mathrm{R}/\mathrm{A}}_0 (\omega) \,&=\, \frac{1}{\omega - \dunderline{\varepsilon} - \dunderline{\Sigma}^{\mathrm{R}/\mathrm{A}}_\mathrm{res}} \quad, \\
\label{eq:keldysh_greens_function_noninteracting}
\dunderline{G}^{\mathrm{K}}_0 (\omega) \,&=\, \dunderline{G}^{\mathrm{R}}_0 (\omega) \dunderline{\Sigma}^{\mathrm{K}}_\text{res} (\omega) \dunderline{G}^{\mathrm{A}}_0 (\omega) \quad.
\end{align}

The repulsive Coulomb interaction between the dot electrons leads to a renormalization of the self-energy. Here, we compute this renormalization using the fRG approach. This yields an RG equation for the self-energy, which can be expressed diagrammatically as\cite{frg_review}
\begin{align}
\label{eq:selfenergy_rg_equation_diagrammatic}
\begin{tikzpicture}[baseline=-0.6ex,scale=0.4] 
\coordinate (left) at (-0.9,0.0) {};
\coordinate (right) at (0.9,0.0) {};
\node[selfenergy] (middle) at (0.0,0.0) {$\Sigma$};
\node[derivative] (above) at (0.0,1.5) {};
\draw[line1] (right) -- (middle);
\draw[line1] (middle) -- (left);
\end{tikzpicture}
\,&=\,
\begin{tikzpicture}[baseline=-0.6ex,scale=0.4]
\coordinate (left) at (-0.9,0.0) {};
\coordinate (right) at (0.9,0.0) {};
\node[selfenergy] (middle) at (0.0,0.0) {$\gamma_2$};
\coordinate (above) at (0.0,2.0) {};
\path[singlescale] (middle) to[left, loop, out=135, in=45,looseness=7] (middle);
\draw[line1] (right) -- (middle);
\draw[line1] (middle) -- (left);
\end{tikzpicture} \quad.
\end{align}
The diagram on the left-hand side represents the derivative of the self-energy with respect to the flow parameter $\Lambda$, while the diagram on the right-hand side is of Hartree-Fock form in Hugenholtz representation. Here, the single-scale propagator (crossed line) replaces the free contraction line and the interaction vertex represents the two-particle vertex function $\gamma_2(\Lambda)$.

In general, the two-particle vertex function $\gamma_2(\Lambda)$ can be obtained from a corresponding RG equation within the fRG approach. The right hand side of the RG equations for the $n$-particle vertex $\gamma_n(\Lambda)$ with $n \geq 2$ depends on $\gamma_{n+1}(\Lambda)$. This leads to a hierarchy of infinitely many RG equations \cite{frg_review}. Here, we disregard all vertex corrections and insert the bare vertex
\begin{align}
\label{eq:vertex_frg}
\overline{v}_{l_1 l_2, l^\prime_1 l^\prime_2} \,=\, \begin{cases}
U & \text{if} \quad l_1=l^\prime_1 \neq l_2 = l^\prime_2 \\
-U & \text{if} \quad l_1 = l^\prime_2 \neq l_2 = l^\prime_1 \\
0 & \text{else}
\end{cases}
\end{align}
for $\gamma_2(\Lambda)$. This means a truncation of the hierarchy of RG equations in lowest order. It corresponds to an RG-enhanced perturbation theory to leading order in $U$. Translating the diagram in \eqref{eq:selfenergy_rg_equation_diagrammatic} es explained in Ref.~\onlinecite{jakobs_meden_schoeller_2007} yields
\begin{align}
\label{eq:retarted_selfenergy_rq_equation_starting_point}
\td{}{\Lambda} \Sigma^\mathrm{R}_{ll^\prime}(\Lambda) \,&=\, - \frac{i}{4 \pi} \sum_{l_1 l_1^\prime} \overline{v}_{l l_1, l^\prime l^\prime_1}  \int \mathrm{d} \omega \, S_{l^\prime_1 l_1}^\mathrm{K}(\Lambda,\omega) \quad,\\
\notag
\td{}{\Lambda} \Sigma^\mathrm{K}_{ll^\prime}(\Lambda) \,&=\, - \frac{i}{4 \pi} \sum_{l_1 l_1^\prime} \overline{v}_{l l_1, l^\prime l^\prime_1} \int \mathrm{d} \omega \, \left[\dunderline{S}^\mathrm{R}(\Lambda,\omega)  \right. \\
\label{eq:keldysh_selfenergy_rq_equation_starting_point}
& \quad \left. - \dunderline{S}^\mathrm{A}(\Lambda,\omega) \right]_{l^\prime_1 l_1} \quad,
\end{align}
where $\dunderline{S}^x(\Lambda,\omega)$ denotes the three components ($x=\mathrm{R},\mathrm{A},\mathrm{K}$) of the single-scale propagator, which is defined as
\begin{align}
\notag
\dunderline{S} (\Lambda,\omega) \,&=\, \left(\begin{array}{cc}
\dunderline{S}^{\mathrm{R}} (\Lambda,\omega) & \dunderline{S}^{\mathrm{K}} (\Lambda,\omega) \\
0 & \dunderline{S}^{\mathrm{A}} (\Lambda,\omega)
\end{array} \right) \\
\label{eq:singlescale_propagator}
&=\,- \dunderline{G} (\Lambda,\omega)\, \bigg\lbrace\td{}{\Lambda} \Big[\dunderline{G}_0 (\Lambda,\omega)\Big]^{-1}\bigg\rbrace\,\dunderline{G} (\Lambda,\omega) \quad.
\end{align}

The $\Lambda$-dependence of the Green's and vertex functions is established by supplementing an infrared cut-off $\Lambda$ to the Green's function. It allows to treat the energy scales of the system successively from high to low energies. Starting from $\Lambda\,=\,\infty$, where the free propagation is completely suppressed, the fRG describes the scaling of the effective vertices and the self-energy during the process of successively turning on the free propagation of the model by reducing $\Lambda$. This means that the RG equations are solved along the RG path from $\Lambda\,=\, \infty$ to $\Lambda\,=\,0$, where the original problem is recovered. Technically, this approach constitutes a means to resum systematically certain classes of diagrams in the perturbative series representation of the self-energy.

A crucial step is therefore to introduce an appropriate cut-off in the Green's function. The hybridization flow\cite{jakobs_pletyukhov_schoeller_2010b} has proved to be a convenient choice in nonequilibrium since it preserves fundamental symmetries as the Kubo-Martin-Schwinger conditions and causality. Physically, the idea is to couple the quantum dot uniformly to an auxiliary reservoir. This results in an additional addend to the self-energy of the form
 \begin{align}
\label{eq:retarded_auxiliary_reservoir_selfenergy}
\left(\dunderline{\Sigma}^\mathrm{R/A}_\mathrm{aux} (\Lambda) \right)_{ll^\prime} \,&=\, \mp i \delta_{ll^\prime} \Lambda \quad, \\
\label{eq:keldysh_auxiliary_reservoir_selfenergy}
\left(\dunderline{\Sigma}^\mathrm{K}_\mathrm{aux} (\Lambda) \right)_{ll^\prime} \,&=\, -2 i \delta_{ll^\prime} \left[1 - 2 f_\mathrm{aux}(\omega - \mu_\mathrm{aux}) \right] \Lambda \quad,
\end{align}
while the hybridization $\Lambda$ serves as the cut-off. We assume $T_\mathrm{aux}=\infty$ which leads to $f_\mathrm{aux}(\omega - \mu_\mathrm{aux})={1 \over 2}$, i.e., a flat distribution, with the consequence that the contribution to the Keldysh component vanishes, $\Sigma^\mathrm{K}_\mathrm{aux} (\Lambda) \,=\,0$. This prevents the auxiliary reservoir from implying an additional structure like, e.g., Fermi edges, to the theoretical description of the nonequilibrium stationary state.  Furthermore, the single-scale propagator becomes \cite{jakobs_pletyukhov_schoeller_2010a,jakobs_pletyukhov_schoeller_2010b}
\begin{align}
\label{eq:retarded_singlescale_propagator}
\dunderline{S}^{\mathrm{R}/\mathrm{A}}(\Lambda) \,&=\, \mp i \dunderline{G}^{\mathrm{R}/\mathrm{A}}(\Lambda) \dunderline{G}^{\mathrm{R}/\mathrm{A}}(\Lambda) \quad,\\ 
\label{eq:keldysh_singlescale_propagator}
\dunderline{S}^\mathrm{K}(\Lambda) \,&=\, -i \dunderline{G}^\mathrm{R}(\Lambda) \dunderline{G}^\mathrm{K}(\Lambda) + i \dunderline{G}^\mathrm{K}(\Lambda) \dunderline{G}^\mathrm{A}(\Lambda)
\end{align}
with
\begin{align}
\label{eq:retarded_greens_function_rg}
\dunderline{G}^{\mathrm{R}} (\Lambda,\omega) \,&=\, \frac{1}{i \Lambda + \omega - \dunderline{\Sigma}^{\mathrm{R}} (\Lambda,\omega)} \,=\, [\dunderline{G}^{\mathrm{A}}(\Lambda,\omega)]^\dagger \quad,
\end{align}
and the Keldysh component follows from the relation \eqref{eq:keldysh_greens_function} which holds also for the $\Lambda$-dependent Green's function. Here, we have separated the auxiliary reservoir contribution \eqref{eq:retarded_auxiliary_reservoir_selfenergy} from the self-energy $\dunderline{\Sigma}^{\mathrm{R}} (\Lambda,\omega)$.

Solving the RG equations for the self-energy requires their initial conditions at the starting point $\Lambda\,=\,\infty$. Setting them up as explained in Ref.~\onlinecite{jakobs_pletyukhov_schoeller_2010a} gives
\begin{align}
\notag
\Sigma^{\mathrm{R}}_{ll^\prime} (\Lambda,\omega) \bigg|_{\Lambda=\infty} \,&=\, \varepsilon_{ll^\prime} + \frac{1}{2} \sum_{l_1} \overline{v}_{ll_1,l^\prime l_1} - \frac{i}{2} \Gamma_{ll^\prime} \\
\label{eq:retarded_selfenergy_initial_conditions}
&=\, \left( h_{l} - V_{g} \right) \delta_{ll^\prime} - \frac{i}{2} \Gamma_{ll^\prime} \quad, \\
\label{eq:keldysh_selfenergy_initial_conditions}
\Sigma^{\mathrm{K}}_{ll^\prime} (\Lambda,\omega) \bigg|_{\Lambda=\infty} \,&=\, - i \sum_\alpha  \Gamma^\alpha_{ll^\prime} \sgn(\omega - \mu_\alpha) \quad,
\end{align}
where the second term on the right hand side of the first line in \eqref{eq:retarded_selfenergy_initial_conditions} is the contribution from the Hartree diagram, which at $\Lambda\,=\,\infty$ is the only non-vanishing correction from the diagrammatic series representation of the self-energy.

The retarded (advanced) component of the Green's function as a function of the frequency $\omega$ is analytic in the upper (lower) half of the complex plane. This together with the frequency-independence of the (bare) vertex has the important consequence that the integral on the right hand side of \eqref{eq:keldysh_selfenergy_rq_equation_starting_point} vanishes. This yields
\begin{align}
\label{eq:keldysh_selfenergy_rg_equation}
\td{}{\Lambda} \dunderline{\Sigma}^\text{K}(\Lambda) \,=\, 0 \quad,  
\end{align}
i.e. the Keldysh component of the self-energy does not renormalize.

In contrast, the frequency integral on the right hand side of \eqref{eq:retarted_selfenergy_rq_equation_starting_point} is non-vanishing and can be evaluated analytically using the spectral representation of the retarded component of the self-energy. This is possible since the (bare) two-particle interaction vertex is independent of the frequency. The resulting expressions can be found in appendix \ref{app:selfenergy}. As a result, the right hand side of the RG equation \eqref{eq:retarted_selfenergy_rq_equation_starting_point} is a self-adjoint matrix since $(\dunderline{S}^\mathrm{K}(\Lambda,\omega))^\dagger\,=-\,\dunderline{S}^\mathrm{K}(\Lambda,\omega)$. Thus, we obtain a renormalized dot Hamiltonian
\begin{align}
\label{eq:H_0_renormalized}
\tilde{H}_0  \,=\, \sum_{ll^\prime}\, \tilde{\varepsilon}_{ll^\prime}\, c_{l}^{\dagger}\, c_{l^\prime} \quad,
\end{align}
with $\dunderline{\tilde{\varepsilon}} \,=\, \dunderline{\Sigma}^{\mathrm{R}} (\Lambda=0) - \dunderline{\Sigma}^\mathrm{R}_\text{res}$ for $\Lambda=0$. The reservoir dressed Green's function is therefore the one of a non-interacting \textit{open} system with
\begin{align}
\label{eq:renormalized_retarded_greens_function}
\dunderline{G}^{\mathrm{R}/\mathrm{A}} (\omega) \,&=\, \frac{1}{\omega - \dunderline{\tilde{\varepsilon}} - \dunderline{\Sigma}^{\mathrm{R}/\mathrm{A}}_\mathrm{res}} \quad, \\
\label{eq:renormalized_keldysh_greens_function}
\dunderline{G}^{\mathrm{K}} (\omega) \,&=\, \dunderline{G}^{\mathrm{R}} (\omega) \dunderline{\Sigma}^{\mathrm{K}}_\text{res} (\omega) \dunderline{G}^{\mathrm{A}} (\omega) \quad.
\end{align}
This has the consequence that, as is shown in Ref.~\onlinecite{meir_wingreen_92}, the current formula \eqref{eq:current_frg} reduces to  the Landauer-B\"uttiker formula
\begin{align}
\label{eq:landauer_buettiker}
I^\mathrm{st}_\gamma \,=\, \frac{1}{2 \pi}\sum_\alpha \int \mathrm{d} \omega \, T_{\gamma \alpha} (\omega) \big[f_\gamma(\omega - \mu_\gamma) - f_\alpha(\omega - \mu_\alpha)\big] \quad,
\end{align}
where
\begin{align}
\label{eq:transmission_probability}
T_{\gamma \alpha} (\omega) \,=\, \tr \dunderline{\Gamma}^\gamma \dunderline{G}^\mathrm{R} (\omega) \dunderline{\Gamma}^\alpha \dunderline{G}^\mathrm{A} (\omega)
\end{align}
is the transmission probability.

To summarize, the fRG approach in lowest order truncation is a means to compute the static self-energy with the effect of a renormalization of the single-particle dot Hamiltonian.

\subsection{Numerical RG}
\label{subsec:nrg}
We benchmark the solutions obtained from the RTRG and the fRG approach, each constituting a \textit{perturbative} RG method, against highly accurate NRG data in the linear response regime. To obtain the most accurate NRG data, we restrict the model to the case of proportional coupling, i.e. $\dunderline{\Gamma}^\alpha\,=\, x_\alpha \dunderline{\Gamma}$ with $\sum_\alpha x_\alpha\,=\,1$. In this case, as shown in Ref.~\onlinecite{meir_wingreen_92}, one can again use the Landauer-B\"uttiker-type formula \eqref{eq:landauer_buettiker} but with the transmission probability \eqref{eq:transmission_probability} given by 
\begin{align}
\label{eq:transmission_probability_proportional}
T_{\gamma \alpha}(\omega) \,=\, 2 \pi\, x_\gamma\, x_\alpha\, \tr \dunderline{\Gamma}\, \dunderline{\rho}(\omega) \quad,
\end{align}
where $\dunderline{\rho}(\omega)\,=\, {i \over 2 \pi} (\dunderline{G}^\mathrm{R} - \dunderline{G}^\mathrm{A})(\omega)$ is the dot spectral function. This quantity characterizes completely the current across the dot in linear response. To see this, we first note that $f_\gamma(\omega - \mu_\gamma) - f_\alpha(\omega - \mu_\alpha) \approx - f^\prime(\omega) (\mu_\gamma-\mu_\alpha)$. As a consequence, with $\mu_\alpha\,=\,-e V_\alpha$, the current is recast as
\begin{align}
\label{eq:current_linear_response}
I_\gamma \,=\, \sum_\alpha\, G_{\gamma \alpha}\, (V_\gamma - V_\alpha) \quad,
\end{align} 
with the conductance tensor
\begin{align}
\label{eq:conductance_tensor}
G_{\gamma \alpha} \,=\, - \frac{1}{2 \pi} \int\mathrm{d}\omega\, T_{\gamma \alpha}(\omega)\, f^\prime(\omega)  \,=\, \frac{1}{2 \pi} T_{\gamma \alpha}(0) \quad,
\end{align}
where we used $f^\prime(\omega)\,=\,-\,\delta(\omega)$ in the last step.

The calculations are performed using the full-density-matrix NRG \cite{Weichselbaum2007-2012}
and make use of the QSpace tensor library developed by A. Weichselbaum \cite{Weichselbaum2012a}.
We employ an efficient, interleaved NRG setup \cite{Stadler2016} with an overall discretization parameter of 
$\Lambda=6$ (i.e., $\sqrt[3]{6}$ between each truncation), and we keep states up to a
rescaled energy of $E_{\textrm{trunc}}=10$ and maximal number $N_{\textrm{keep}}=4000$ 
during the NRG iteration. Additionally, results are averaged between two realizations of the discretization 
($z$ averaging \cite{Bulla2008,Oliveira1994}). The wide-band and zero-temperature limit are practically realized 
by setting the half-bandwidth to $10^4$ and temperature to $10^{-8}$. We checked
that our results are converged up to the percent level with respect to all involved numerical
parameters. Finally, we note that one need not broaden the NRG data as
the linear conductance can be inferred from discrete spectral weights.

\section{Conductance in the linear response regime}
\label{sec:linear_response}
In order to demonstrate the strength of the RTRG method in describing charge fluctuations, we discuss results for a generic quantum dot with three ($Z=3$) non-degenerate  levels, i.e., $\abs{h_l - h_{l^\prime}} \sim \Gamma$ , and two reservoirs held at different chemical potential and zero temperature. The difference between the chemical potentials of the reservoirs is quantified by the bias voltage, i.e. $\mu_L - \mu_R\,=\,V$ and $\mu_{L/R} \,=\,\pm {V \over 2}$. In particular, we consider the first derivative of the current $I_\gamma$, the differential conductance
\begin{align}
\label{eq:conductance}
G \,=\, G_{LR} \,=\, \td{}{V} I_L \,=\, - \td{}{V} I_R \quad.
\end{align} 
As a first step, we benchmark the RTRG (and the fRG) method in the chosen approximation against NRG data. We can do this for an arbitrary model with proportional coupling in the linear response regime.

We parameterize the tunneling matrix $t^\alpha_{l l^\prime}$ of the model as
\begin{align}
\label{eq:tunneling_matrix_parameterization}
t^\alpha_{l l^\prime} \,=\, \sqrt{\frac{\overline{\Gamma}_{\alpha l l^\prime}}{2 \pi}} \mathrm{e}^{i \overline{\varphi}_{\alpha l l^\prime} \pi} \quad,
\end{align}
leading to
\begin{align}
\label{eq:hybridization_matrix_parameterization}
\Gamma^\alpha_{ll^\prime} \,=\, \sum_{l_1} \sqrt{\overline{\Gamma}_{\alpha l_1 l} \overline{\Gamma}_{\alpha l_1 l^\prime}} e^{-i(\overline{\varphi}_{\alpha l_1 l} -\overline{\varphi}_{\alpha l_1 l^\prime}) \pi} 
\end{align}
for the hybridization matrix. In the case of proportional coupling, we introduce the ratio $\kappa \,=\, \overline{\Gamma}_{R l l^\prime}/\overline{\Gamma}_{L l l^\prime}$.

We consider an arbitrary hybridization matrix. To this end, we present here the results for a model with hybridization matrix parameterized by random numbers. Table~\ref{tab:proportional_coupling} contains the corresponding parameters $\overline{\Gamma}_{\alpha l l^\prime}$ and $\overline{\varphi}_{\alpha l l^\prime}$.

\begin{table}
\begin{tabular}{|c|c|}
\hline
$(\overline{\Gamma}_{\text{L}11},\overline{\varphi}_{\text{L}11})$ & $(0.00680672,0.98)$ \\ 
$(\overline{\Gamma}_{\text{L}12},\overline{\varphi}_{\text{L}12})$ & $(0.0605042,0.96)$ \\
$(\overline{\Gamma}_{\text{L}13},\overline{\varphi}_{\text{L}13})$ & $(0.0332773,0.12)$ \\
$(\overline{\Gamma}_{\text{L}21},\overline{\varphi}_{\text{L}21})$ & $(0.0627731,-0.99)$ \\
$(\overline{\Gamma}_{\text{L}22},\overline{\varphi}_{\text{L}22})$ & $(0.0589916,0.79)$ \\
$(\overline{\Gamma}_{\text{L}23},\overline{\varphi}_{\text{L}23})$ & $(0.024958,-0.16)$ \\
$(\overline{\Gamma}_{\text{L}31},\overline{\varphi}_{\text{L}31})$ & $(0.00983193,-0.8)$ \\
$(\overline{\Gamma}_{\text{L}32},\overline{\varphi}_{\text{L}32})$ & $(0.0468908,0.71)$ \\
$(\overline{\Gamma}_{\text{L}33},\overline{\varphi}_{\text{L}33})$ & $(0.0559664,0.8)$ \\
\hline
$\kappa$ & 1.77778 \\
\hline
\end{tabular}
\caption{Input parameters for the tunneling matrix $\dunderline{t}^L$ of the left reservoir in the case of proportional coupling. These parameters define the matrix elements $t^L_{l l^\prime}$ via \eqref{eq:tunneling_matrix_parameterization}. The tunneling matrix of the right reservoir follows from the relation $\dunderline{t}^R \,=\, \sqrt{\kappa} \dunderline{t}^L$.}
\label{tab:proportional_coupling}
\end{table}

\begin{figure}
\subfloat[$U/\Gamma=0.1$\label{fig:linear_conductance_U_0.1}]{\includegraphics[scale=1.0]{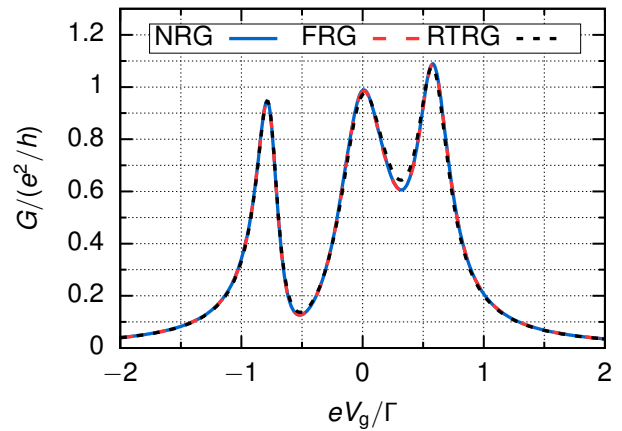}}

\subfloat[$U/\Gamma=1.0$\label{fig:linear_conductance_U_1.0}]{\includegraphics[scale=1.0]{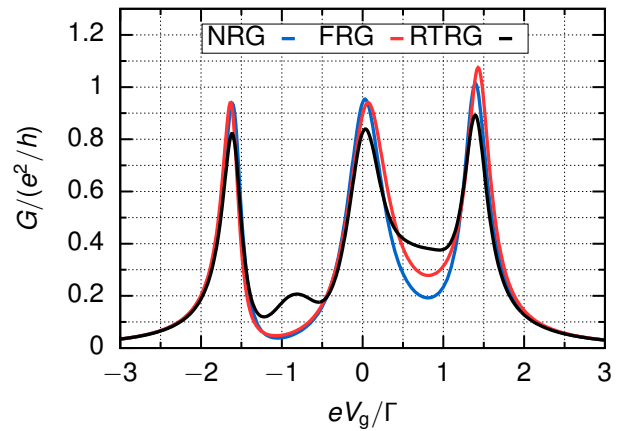}}
\caption{Linear conductance $G$ as function of the gate voltage for the model with $\dunderline{t}^\alpha$ defined by Table~\ref{tab:proportional_coupling}, and level-spacings $h_l/\Gamma=(-0.7,0.0,0.5)$. We set $D=1000.0\Gamma$ for all numerical RTRG calculations in this article. All three applied methods (NRG, fRG and RTRG) are in agreement regarding the position and shape of the conductance peaks.\label{fig:linear_conductance}}
\end{figure}

\begin{figure*}
\centering
\includegraphics[scale=1.0]{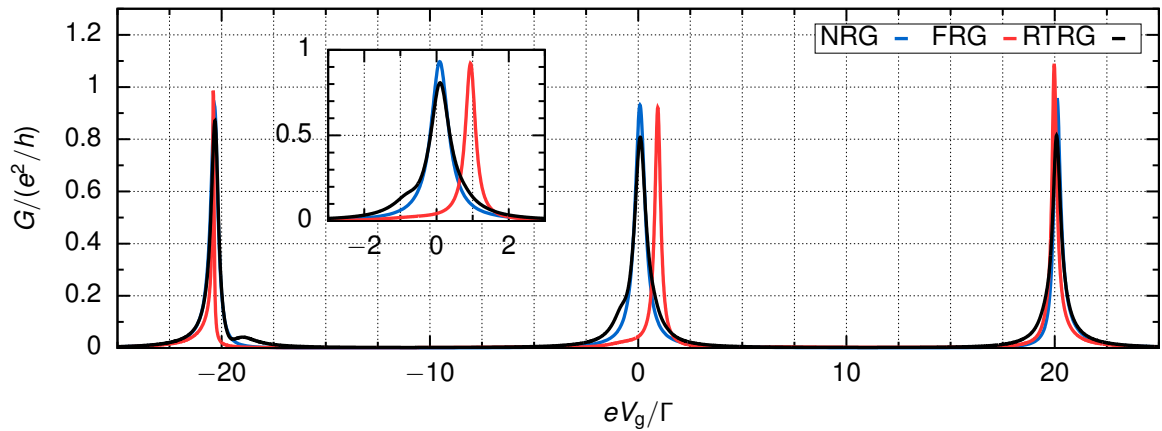}
\caption{Linear conductance $G$ as function of the gate voltage $V_\text{g}$ for the model with $\dunderline{t}^\alpha$ defined by Table~\ref{tab:proportional_coupling}, $U=20.0\Gamma$ and level-spacings $h_l/\Gamma=(-0.7,0.0,0.5)$. The inset shows a close-up of the central peak, clearly revealing the deviations in position and shape of the maximum within the fRG solution. In contrast, NRG and RTRG data are in good agreement regarding the position and the width of the condutance peak.\label{fig:linear_conductance_U_20.0}}
\end{figure*}

We found for arbitrary strengths of the Coulomb interaction three peaks for the conductance $G$ as a function of the gate voltage $V_\text{g}$. Figs. \ref{fig:linear_conductance} and \ref{fig:linear_conductance_U_20.0} show exemplary results. This outcome is commonly interpreted using the picture of the Coulomb blockade, see e.g. Ref.~\onlinecite{review_qi} for a review. Accordingly, a peak occurs whenever the ground states of the $N$ and $N+1$ particle sectors are degenerate and a resonant electron transport across the quantum dot is possible. This is the meaning of the condition \eqref{eq:resonant_tunneling_rtrg} for resonant tunneling, which differs from \eqref{eq:resonant_tunneling_pt} only due to the renormalization of the peak positions. In contrast, the conductance is drastically reduced between the peaks, resulting in so-called \textit{Coulomb blockade valleys}. The dot electron number is fixed in this case and tunneling in and out of the dot involves the occupation of a dot state with a different particle number. These states are of higher energy and the occupation of these states become more and more suppressed for an increasing Coulomb interaction. Correspondingly, the Coulomb blockade valleys are more pronounced for increasing $U/\Gamma$.  

The Green's function formalism provides an alternative interpretation. In this case, we deduce from \eqref{eq:transmission_probability_proportional} and \eqref{eq:conductance_tensor} that the peaks in the conductance $G$ are the maxima of the dot spectral function
\begin{align}
\label{eq:dot_spectral_function}
\dunderline{\rho} (0) \,=\, \frac{1}{\pi} \sum_k \imag \Bigg\lbrace\frac{1}{\tilde{h}_k - V_\text{g} - i \frac{\tilde{\Gamma}_k}{2}} \dunderline{\tilde{P}}_k \Bigg\rbrace \quad.
\end{align}
Here, we inserted the spectral decomposition of $\dunderline{\Sigma}^{\text{R}} (\Lambda=0) \,=\, \sum_k \tilde{\lambda}_k \dunderline{\tilde{P}}_k$ where $\tilde{\lambda}_k \,=\, \tilde{h}_k - V_\text{g} - i \frac{\tilde{\Gamma}_k}{2}$ are the eigenvalues and $\dunderline{\tilde{P}}_k$ the corresponding projector. $\tilde{h}_k$ has the meaning of the position of a renormalized single-particle energy while $\tilde{\Gamma}_k$ is the corresponding level broadening. Due to \eqref{eq:dot_spectral_function}, a conductance peak occurs for $\tilde{h}_k\,=\,V_\text{g}$, i.e. resonant tunneling is obtained when the gate voltage equals a single-particle energy. Simultaneously, the very same level being unoccupied for $V_\text{g} < \tilde{h}_k$ becomes populated with one electron at this point. In conclusion, the fRG solution in lowest order truncation scheme complies with an effective single-particle picture for the three conductance peaks occurring in the linear response regime.

We find a very good agreement between all three considered methods in the regime of \textit{small interaction strengths} $U \ll  \Gamma$. For example, Fig.~\ref{fig:linear_conductance_U_0.1} shows the linear conductance as function of the gate voltage for $U=0.1\Gamma$. While an agreement between fRG and NRG was expected in this regime, the RTRG data for the conductance is also reliable, as it was already noted for single-level\cite{schoeller_1999} and for two-level quantum dots\cite{boese_hofstetter_schoeller_2001,saptsov_wegewijs_2012,schoeller_koenig_2000}.

Fig. \ref{fig:linear_conductance_U_1.0} is exemplary for the solutions from the three methods in the regime of \textit{intermediate interaction strengths}, $U \sim \Gamma$. In this case, we still find a good agreement between fRG, NRG and RTRG data regarding the position and width of the conductance peaks. However, the shape of the RTRG solution deviates from the NRG solution in the Coulomb blockade valleys. These deviations are perceptible imprints of the increasing significance of orbital fluctuations due to cotunneling processes in the quantum dot  for increasing interaction strengths. Fourth-order terms in the tunneling are necessary for a reasonable description of the cotunneling processes. However, these terms are only taken partially into account within the considered truncation scheme for the RTRG approach, as is discussed in appendix \ref{app:perturbation_theory}. Thus, it is no surprise that the RTRG data is less reliable within the Coulomb blockade valleys. This means that the employed approximation for the RTRG equation describes charge fluctuations reliably, but is insufficient to study cotunneling processes. In contrast, these processes are fully taken into account by the fRG approach. The corresponding results thus show a good agreement with the NRG data also in the Coulomb blockade valleys.

Lastly, we considered the regime of large interaction strengths ($U \gg \Gamma$). Fig. \ref{fig:linear_conductance_U_20.0} shows the conductance as the function of the gate voltage for $U=20.0 \Gamma$. In this case, we find again a good accordance between RTRG and NRG data. In contrast, the fRG solution clearly shows deviations from the NRG solution for the position and shape of the conductance peaks. This is most pronounced for the peak arising from the transition from $N=1$ to $N=2$. In this case, the fRG method shifts the position of the peak further away from the particle-hole symmetric point than the other two methods, see the inset of Fig.~\ref{fig:linear_conductance_U_20.0}.

The deviations between the fRG solution and the NRG solutions can be easily understood from the fact that the truncation of the RG equations from the fRG approach is motivated by means of an expansion in the Coulomb interaction. Obviously, this is justified formally only for small interaction strengths $U \ll \Gamma$. It is therefore no surprise that the fRG is not reliable for large interaction strengths $U \gg \Gamma$.

A closer look at Fig. \ref{fig:linear_conductance_U_20.0} reveals that the RTRG produces a small peak close to the left conductance peak (referring to the transition $N=0\rightarrow N=1$) and a small shoulder for the middle conductance peak (referring to the transition $N=1\rightarrow N=2$). Again, these anomalies arise from the neglect of orbital fluctuations from higher order diagrams, similiar to the occurence of the anomaly between the resonances for the case of intermediate Coulomb interaction strenght, see Fig. \ref{fig:linear_conductance_U_1.0}. These features depend crucially on the choice of the tunneling matrix elements and the level spacings. However, they are very weak for strong Coulomb interaction and not relevant for the position and line shape of the main charge fluctuation resonances. It has to be studied in the future how these anomalies can be eliminated by a minimial extension of the RTRG, similiar to the more refined but considerably more expensive versions of the RTRG used in Refs.~\onlinecite{saptsov_wegewijs_2012,schoeller_koenig_2000}, where vertex renormalizations were taken into account.

In total, the benchmark against the NRG data for a model with proportional coupling and non-degenerate dot levels in the linear response regime shows that the RTRG method yields reliable results for position and the width of the peaks of the linear conductance for arbitrary dot-reservoir couplings.

\section{Stationary state current in nonequilibrium}
\label{sec:nonequilibrium}
We now turn to a generic quantum dot coupled to two reservoirs with arbitrary values of the bias $V$. This means that the restriction of proportional coupling is lifted in the following. The parameters defining the tunneling matrix and the hybridization matrix, respectively, can be read off from Table~\ref{tab:generic_coupling}.

\begin{table}
\centering
\begin{tabular}{|c|c|c|}
\hline
$\alpha$ & L & R  \\
\hline
\hline
$(\overline{\Gamma}_{\alpha 11},\overline{\varphi}_{\alpha 11})$ & $(0.0434783,-0.8)$ & $(0.101831,-0.88)$ \\ 
$(\overline{\Gamma}_{\alpha 12},\overline{\varphi}_{\alpha 12})$ & $(0.0640732,-0.19)$ & $(0.01373,0.32)$  \\
$(\overline{\Gamma}_{\alpha 13},\overline{\varphi}_{\alpha 13})$ & $(0.0743707,0.71)$ & $(0.0789474,-0.64)$ \\
$(\overline{\Gamma}_{\alpha 21},\overline{\varphi}_{\alpha 21})$ & $(0.0446224,0.17)$ & $(0.0480549,-0.72)$ \\
$(\overline{\Gamma}_{\alpha 22},\overline{\varphi}_{\alpha 22})$ & $(0.0663616,-0.83)$ & $(0.0915332,-0.08)$ \\
$(\overline{\Gamma}_{\alpha 23},\overline{\varphi}_{\alpha 23})$ & $(0.00457666,0.45)$ & $(0.0560641,0.41)$  \\
$(\overline{\Gamma}_{\alpha  31},\overline{\varphi}_{\alpha 31})$ &$(0.01373,-0.1)$ & $(0.100686,-0.22)$  \\
$(\overline{\Gamma}_{\alpha 32},\overline{\varphi}_{\alpha 32})$ & $(0.0183066,-0.45)$ & $(0.076659,-0.6)$ \\
$(\overline{\Gamma}_{\alpha 33},\overline{\varphi}_{\alpha 33})$ & $(0.0469108,0.19)$ & $(0.0560641,-0.15)$ \\
\hline
\end{tabular}
\caption{Input parameters for the tunneling matrix $\dunderline{t}^\alpha$ of the generic model. These parameters define the matrix elements $t^\alpha_{l l^\prime}$ via \eqref{eq:tunneling_matrix_parameterization}.}
\label{tab:generic_coupling}
\end{table}

\begin{figure*}
\subfloat[$U/\Gamma=0.1,\,V/\Gamma=0.0$\label{fig:non-linear_conductance_U_0.1_V_0.0}]{\includegraphics[scale=1.0]{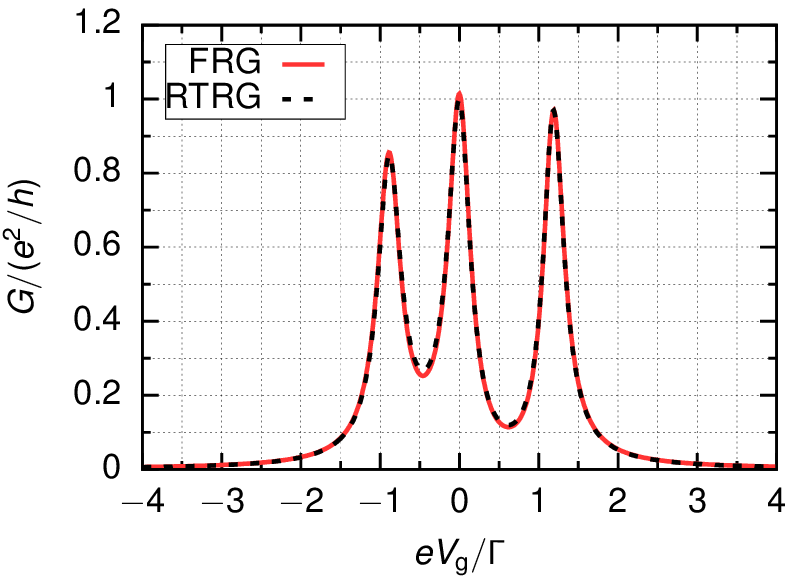}}
\subfloat[$U/\Gamma=1.0,\,V/\Gamma=0.0$\label{fig:non-linear_conductance_U_1.0_V_0.0}]{\includegraphics[scale=1.0]{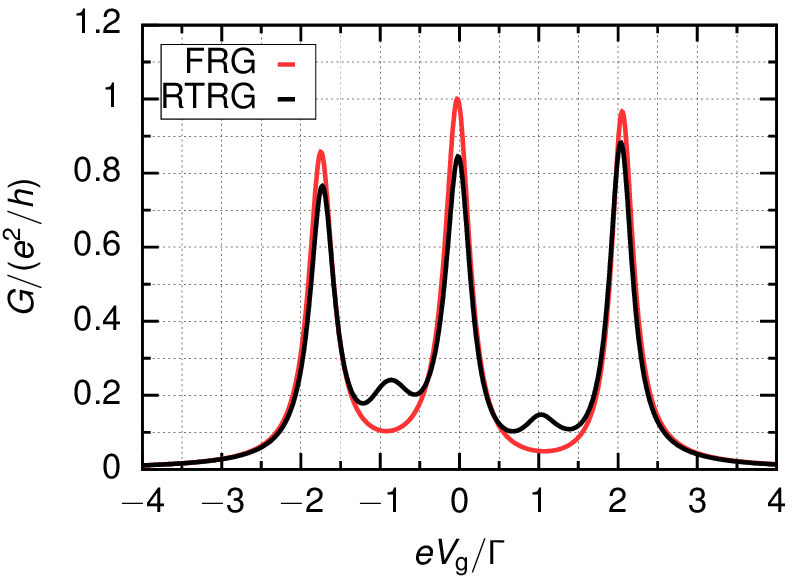}}

\subfloat[$U/\Gamma=0.1,\,V/\Gamma=0.5$\label{fig:non-linear_conductance_U_0.1_V_0.5}]{\includegraphics[scale=1.0]{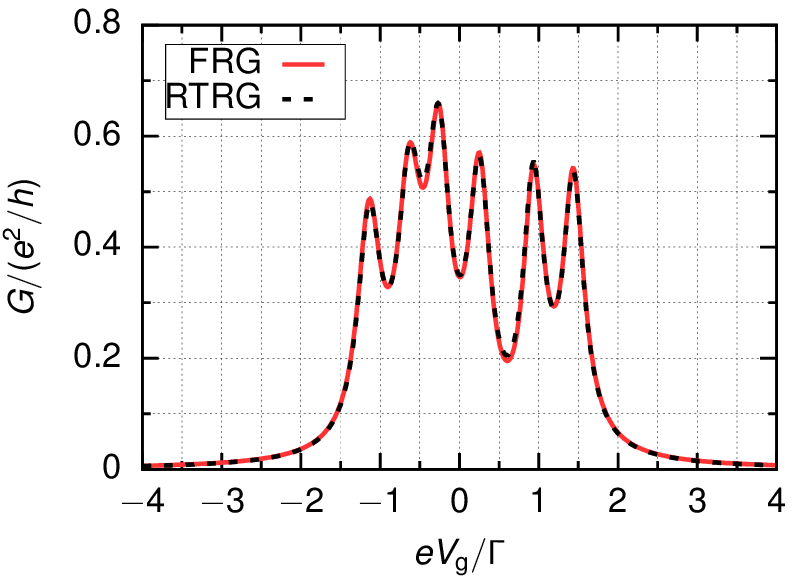}}
\subfloat[$U/\Gamma=1.0,\,V/\Gamma=0.5$\label{fig:non-linear_conductance_U_1.0_V_0.5}]{\includegraphics[scale=1.0]{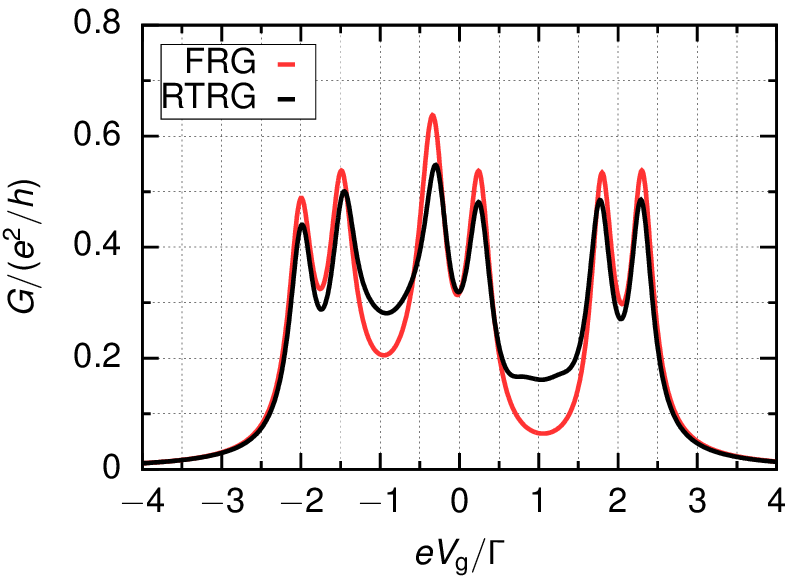}}

\subfloat[$U/\Gamma=0.1,\,V/\Gamma=2.0$\label{fig:non-linear_conductance_U_0.1_V_2.0}]{\includegraphics[scale=1.0]{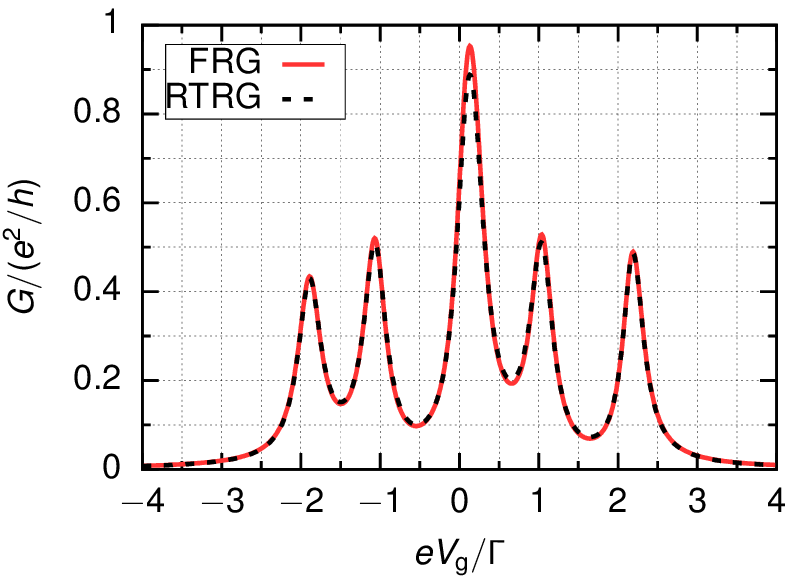}}
\subfloat[$U/\Gamma=1.0,\,V/\Gamma=2.0$\label{fig:non-linear_conductance_U_1.0_V_2.0}]{\includegraphics[scale=1.0]{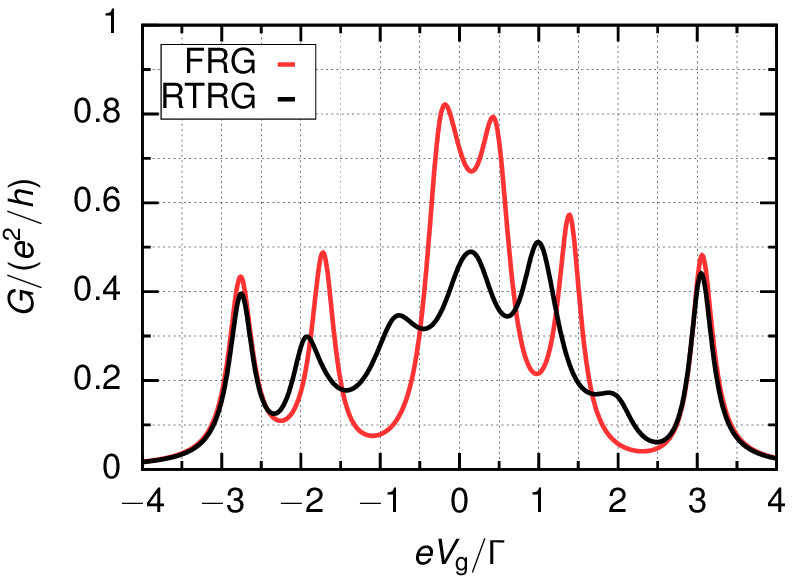}}
\caption{Conductance $G$ as function of the gate voltage $V_\text{g}$ for a model with tunneling matrix $\dunderline{t}^\alpha$ defined in Table~\ref{tab:generic_coupling} and $h_l/\Gamma=(-0.8,0.0,1.1)$ for small to intermediate Coulomb interactions, i.e. $U=0.1\Gamma$ (left panel) and $U=1.0\Gamma$ (right panel).\label{fig:non-linear_conductance}}
\end{figure*}

The fRG approach is controlled in the regime of small Coulomb interaction $U \ll \Gamma$ with the consequence that it can be used as a benchmark to test the reliability of the RTRG approximation in this limit. Our numerical study reveals an almost perfect agreement between RTRG and fRG data for arbitrary bias voltages in this regime. Fig.~\ref{fig:non-linear_conductance} shows exemplary the conductance $G$ as function of the gate voltage $V_\text{g}$ for $U\,=\,0.1\Gamma$ and selected values for $V$. This outcome generalizes our findings in the linear response regime, confirming that the RTRG approach yields accurate results for weak Coulomb interactions also in the limit of strong coupling already within the simplest approximation.

For small Coulomb interaction, the effective single-particle picture is valid. The mere effect of the fRG method in the lowest order truncation scheme is a renormalization of the single-particle dot energy levels $\tilde{h}_k$. Resonant electron transport, causing the conductance peaks, occurs if one of these levels align with the chemical potential of one of the two reservoirs. As a consequence, the conductance peaks are now located at $V_\text{g}=\tilde{h}_k \pm {V \over 2}$. This means that each of the three peaks observed in equilibrium split into two peaks for increasing bias voltage. Eventually, the conductance shows six peaks constituting two groups of three peaks centered at $V_\text{g}=\tilde{h}_2 \pm {V\over2}$ for large bias voltages $V \gg \Gamma$. There is a crossover between the cases of three and six resonances where the number of distinguishable peaks can be smaller than six. This is the case if the distance between two resonance lines is smaller than the peak widths. 

In equilibrium, it is well-established that the fRG yields reliable results from weak to intermediate Coulomb interactions\cite{frg-eq}. However, for large bias voltages the effective single-particle picture is only applicable for small Coulomb interactions. Thus, we cannot use the static fRG data as a benchmark against the RTRG data beyond $U \ll \Gamma$. Nonetheless, we also compared the results for the differential conductance in order to estimate the parameter range where the solutions from both approaches are in qualitative agreement.

We find a more complex behavior for intermediate interaction strengths. The right panel of Fig.~\ref{fig:non-linear_conductance} shows exemplary the evolution of the differential conductance as function of the gate voltage with increasing bias for $U/\Gamma\,=\,1.0$. Similar to Fig.~\ref{fig:linear_conductance_U_1.0}, Fig.~\ref{fig:non-linear_conductance_U_1.0_V_0.0} reveals a good agreement between fRG and RTRG data for the position and width of the conductance peaks in the linear response regime. A qualitative agreement between results from both approaches is also obtained for $V/\Gamma\,=\,0.5$, c.f. Fig. \ref{fig:non-linear_conductance_U_1.0_V_0.5}, where both approaches predict the same position of the six conductance peaks. This is no longer the case already for moderate bias $V/\Gamma\,=\,2.0$. Fig. \ref{fig:non-linear_conductance_U_1.0_V_2.0} shows that in this case the fRG and the RTRG approach agree only for the outer conductance peaks, i.e. the left-most and the right-most peaks. In contrast, the RTRG solution shows an essentially different structure compared to the fRG solution in the region between these two peaks.

\begin{figure}
\subfloat[$U/\Gamma=0.1$\label{fig:non-linear_conductance_U_0.1_V_5.0}]{\includegraphics[scale=1.0]{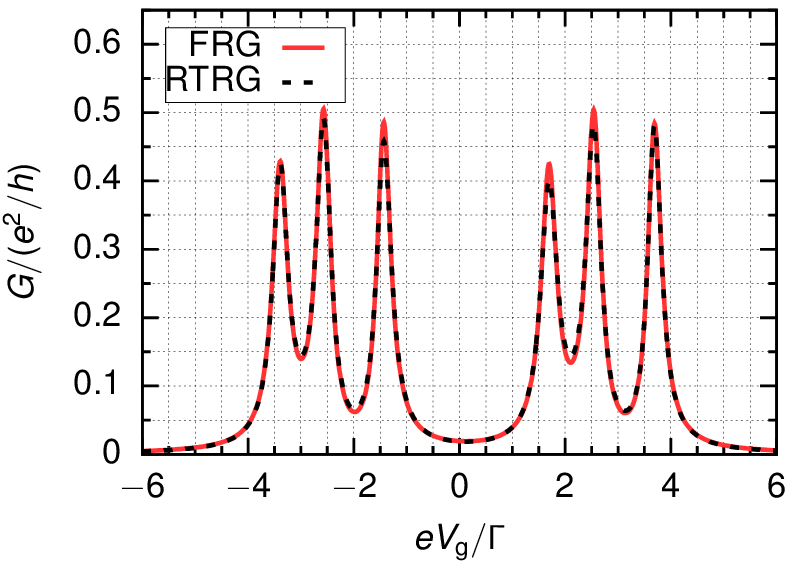}}

\subfloat[$U/\Gamma=0.5$\label{fig:non-linear_conductance_U_0.5_V_5.0}]{\includegraphics[scale=1.0]{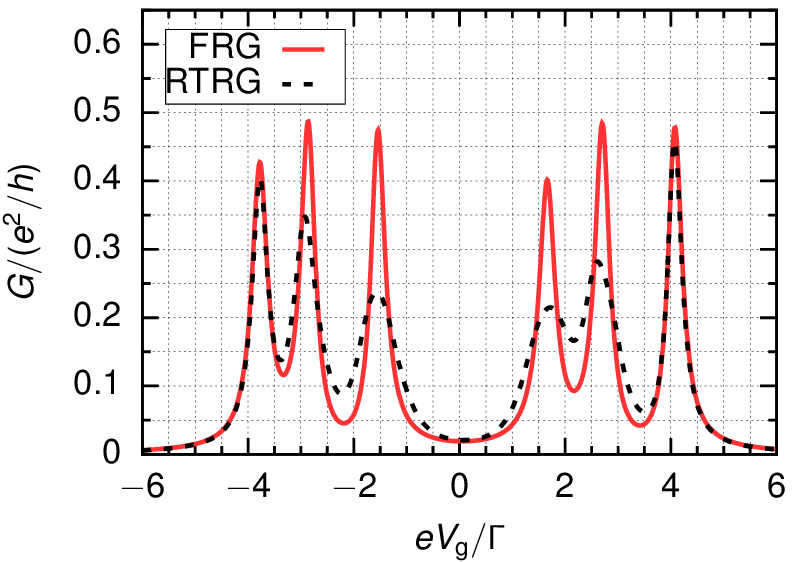}}

\subfloat[$U/\Gamma=2.0$\label{fig:non-linear_conductance_U_2.0_V_5.0}]{\includegraphics[scale=1.0]{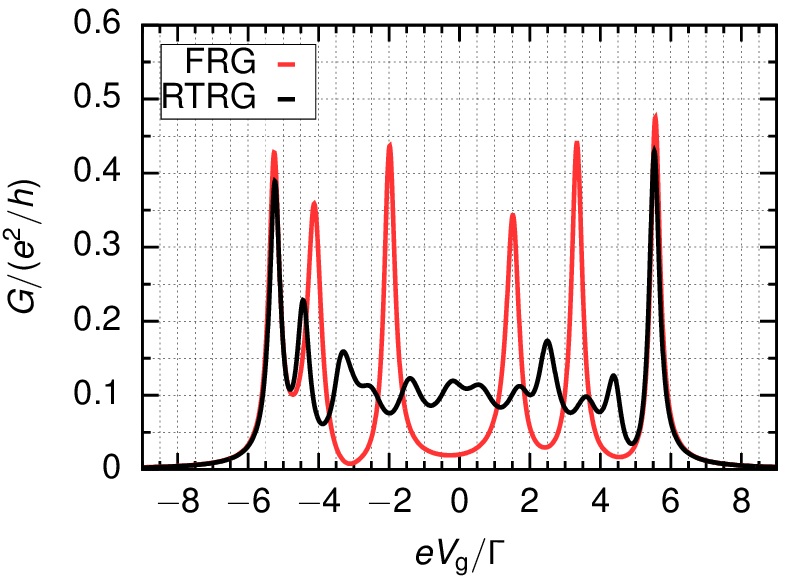}}
\caption{Conductance $G$ as function of the gate voltage $V_\mathrm{g}$ for a model with tunneling matrix $\dunderline{t}^\alpha$ defined in Table~\ref{tab:generic_coupling}, $h_l/\Gamma=(-0.8,0.0,1.1)$ and $V=5.0\Gamma$. While there is a very good agreement between fRG and RTRG solution for small Coulomb interactions $U\,=\,0.1\Gamma$, the results from both approaches coincide only for the outer, i.e. the very left and the very right, peaks for moderate interaction strengths $U\,=\,2.0\Gamma$. In the latter case, the solutions differ significantly in the region between the outer peaks, as explained in the main text.\label{fig:non-linear_conductance_V_5.0}}
\end{figure}

A corresponding picture emerges if we scrutinize the dependence of the differential conductance on the Coulomb interaction at large bias. Fig. \ref{fig:non-linear_conductance_V_5.0} shows the differential conductance as function of the gate voltage for $V/\Gamma \,=\,5.0$ and different values for $U$. Starting from weak coupling ($U/\Gamma\,=\,0.1$, Fig. \ref{fig:non-linear_conductance_U_0.1_V_5.0}), where RTRG and fRG results are in very good agreement, we still find a qualitative agreement for $U/\Gamma\,=\,0.5$, see Fig. \ref{fig:non-linear_conductance_U_0.5_V_5.0}. In particular, both solutions are in accordance regarding the number and position of the conductance peaks but differ in the height of the inner conductance peaks. These are of reduced height in the RTRG solution for the differential conductance compared to the fRG data. In contrast, the solutions for the differential conductance from both approaches no longer comply in the region between the outer peaks for larger Coulomb interactions, as it is shown in Fig. \ref{fig:non-linear_conductance_U_2.0_V_5.0} for $U/\Gamma\,=\,2.0$.

\begin{figure*}
\includegraphics[scale=1.0]{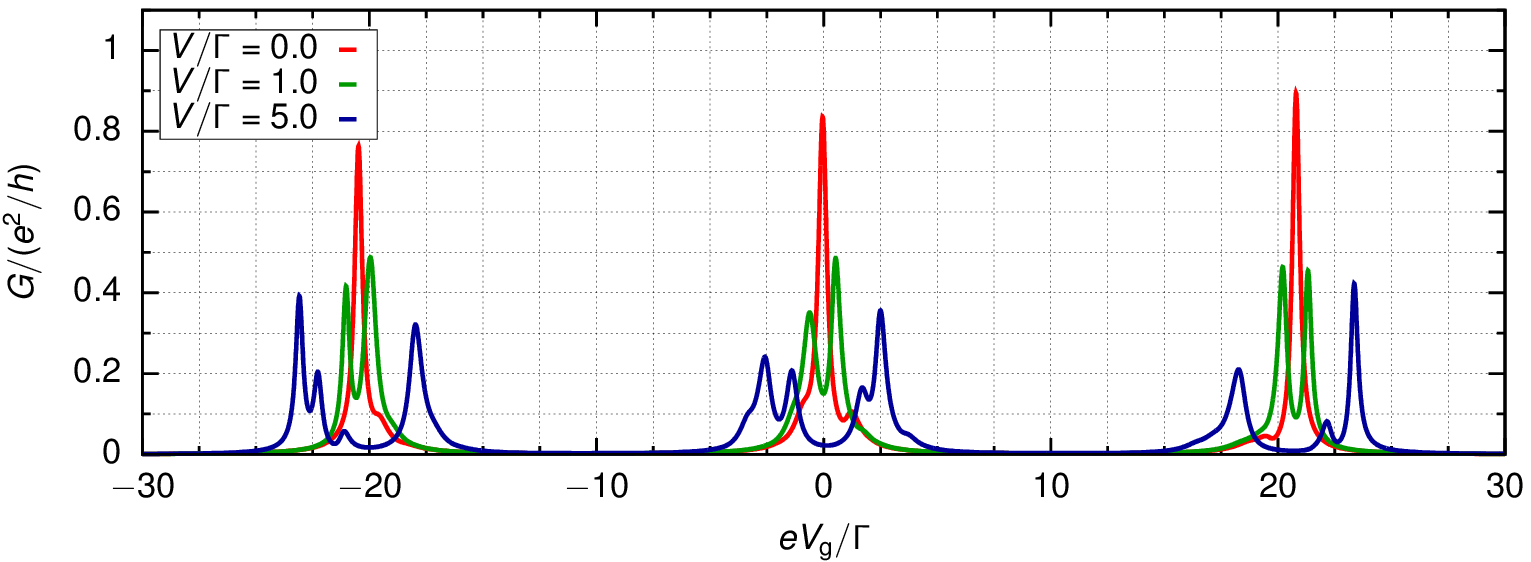}
\caption{RTRG solution for the conductance $G$ as function of the gate voltage $V_\text{g}$ for the model with $\dunderline{t}^\alpha$ defined by Table~\ref{tab:generic_coupling}, $U=20.0\Gamma$, level-spacings $h_l/\Gamma=(-0.8,0.0,1.1)$ and different values for the bias $V$. Each of  the three peaks occurring in the linear response regime ($V\,=\,0$) splits into two peaks of reduced height for increasing bias voltage $V\,=\,\Gamma$. In contrast, additional resonance lines emerge for large enough bias ($V\,=\,5.0\Gamma$).\label{fig:non-linear_conductance_U_20.0}}
\end{figure*}

For intermediate Coulomb interactions and moderate bias, e.g. \ref{fig:non-linear_conductance_U_1.0_V_2.0} and \ref{fig:non-linear_conductance_U_2.0_V_5.0}, the number and positions of the inner conductance peaks is different for the solution from both approaches. In particular, the RTRG solution exhibits more than six local minima which we interpret as additional resonance lines. Their emergence is more pronounced for large Coulomb interaction, as can be seen in Fig. \ref{fig:non-linear_conductance_U_20.0} for $U\,=\,20.0\Gamma$ and $V\,=\,5.0\Gamma$. This behavior of the RTRG solution for the differential conductance can be readily understood from the condition \eqref{eq:resonant_tunneling_rtrg} for resonant tunneling within this approach which is fulfilled if the real part of the eigenvalue $\lambda_k(E)$ of the effective Liouvillian aligns to the chemical potential of one of the two reservoirs. In order to interpret this condition, it is more instructive to consider \eqref{eq:resonant_tunneling_pt}, which determines the resonance lines using perturbation theory. The RG treatment leads to a shift of the resonance lines in the conductance as a function of the gate voltage.

In the linear response regime, i.e., for $V\rightarrow 0$, condition \eqref{eq:resonant_tunneling_pt} is only fulfilled if the ground state energies of the $N$ and $N+1$ electron sectors are degenerate. This means that for $V>0$, one electron can tunnel from the left reservoir onto the dot, occupying the lowest energy many-body state of the $N+1$ electron sector. Afterwards, this electron can leave the dot by tunneling into the right reservoir, resulting in a total tunneling process involving the dot electron numbers $N \rightarrow N+1 \rightarrow N$. As a consequence, the three single-particle dot levels are successively populated with increasing gate voltage $V_\text{g}$. This complies with the single-particle picture and is also the reason why the linear conductance as function of the gate voltage has always three peaks.

If the bias is large enough, \eqref{eq:resonant_tunneling_pt} can also be fulfilled for processes involving excited many-body dot states. For instance, transitions from the ground state of the $N$ particle sector to an excited state of the $N+1$ particle sector can become possible if this condition is matched. Equivalently, these tunneling processes $s_2 \rightarrow s_1$ with $N_{s_1}\,=\,N_{s_2}+1$ are possible if the corresponding energy difference $E_{s_1} - E_{s_2}$ lies within the transport window\cite{review_qi,andergassen_etal}, i.e. $\mu_L > E_{s_1} - E_{s_2} > \mu_R$, provided that the initial state $s_2$ is occupied. As a consequence, additional resonance lines show up in the current, each corresponding to one of these tunneling processes. The emergence of such additional conductance peaks is clearly visible for $U\,=\,20.0\Gamma$ and $V\,=\,5.0\Gamma$ in Fig.~\ref{fig:non-linear_conductance_U_20.0}. We note that each resonance can be split by the bias voltage in at most $4$ resonances. E.g., for the transition $N=0\rightarrow N=1$ (corresponding to the left resonance in Fig.~\ref{fig:non-linear_conductance_U_20.0}), three resonances occur when one of the three renormalized levels matches with the upper chemical potential $\mu_L=V/2$ but only one resonance can appear when the lowest level matches with the lower chemical potential $\mu_R=-V/2$. Once the lowest level is below $\mu_R$ it is occupied and the resonances when the two higher levels match with $\mu_R$ are suppressed by Coulomb blockade. Therefore, for bias voltage significantly larger than $\Gamma$ four resonances are observed in Fig.~\ref{fig:non-linear_conductance_U_20.0} for the left resonance. Similiar considerations hold for the middle and right resonance, but some of the peaks are hardly visible due to broadening effects. Similar findings were reported for an RTRG study of the Anderson model in the regime of strong Coulomb interactions in Ref.~\onlinecite{saptsov_wegewijs_2012}.

One must also distinguish between the deviations observed in the Coulomb blockade valleys in the linear response regime, see Figs. \ref{fig:linear_conductance_U_1.0} and \ref{fig:non-linear_conductance_U_1.0_V_0.0}, and the behavior at intermediate bias $V \sim U \sim \Gamma$. While charge fluctuations are suppressed in the former case, the Coulomb blockade is lifted in the latter case. This means that charge fluctuations are dominant again for $V>U$.  These processes are captured by the RTRG approximation considered in this work. Further evidence that the RTRG solution is reliable in this regime arises from the fact that it yields the exact Liouvillian in the limit $V \rightarrow \infty$. In this case, the right hand side of the RG equation \eqref{eq:liouvillian_rg_real_flow_parameter} is zero which leads to
\begin{align}
\label{eq:liouvillian_infinite_bias}
L(E) \,=\, L^{(0)}\,+\, L^{(1\mathrm{s})} \quad.
\end{align} 
This is an exact result in this limit since all higher-order terms vanish, as will be explained at the end of Appendix \ref{app:perturbation_theory}.

To conclude, we expect a crossover from the effective single-particle behavior of the quantum dot for small Coulomb interactions $U \ll \Gamma$ to a more complex multi-particle situation, exhibiting further resonances, for large Coulomb interactions $U \gg \Gamma$. Fig.~\ref{fig:non-linear_conductance_V_5.0} shows how this crossover sets in for intermediate Coulomb interactions $U \sim \Gamma$ and $V\,=\,5.0 \Gamma$ in the RTRG solution. In contrast, the effective single-particle picture applies for intermediate Coulomb interactions if the bias voltage is smaller than the Coulomb interaction. This is indicated by a qualitative agreement of the RTRG and fRG solutions, see Figs. \ref{fig:non-linear_conductance_U_1.0_V_0.5} and \ref{fig:non-linear_conductance_U_0.5_V_5.0}.

We refrain here from comparing fRG and RTRG results for the conductance in the regime of strong Coulomb interactions $U \gg \Gamma$ since no agreement can be expected, due to the aforementioned reasons. Fig.~\ref{fig:linear_conductance_U_20.0} shows also clearly the deviations from fRG and RTRG data already in linear response in this regime.

In summary, we conclude that the RTRG method yields reliable results for the conductance in nonequilibrium at arbitrary Coulomb interaction, or, equivalently for arbitrary coupling to the reservoirs. From comparing the RTRG solution with fRG results, we estimate that the effective single-particle picture can be employed in nonequilibrium for bias voltages that are smaller than the Coulomb interaction.

\section{Summary}
\label{sec:summary}
In this article, we presented a comparative study of the electron transport through non-degenerate ($\abs{h_l - h_{l^\prime}}  \sim \Gamma$) quantum dots coupled to two reservoirs via generic tunneling matrices in and out of equilibrium. To this end, we applied very basic approximations of the RTRG and fRG methods, where the effective Liouvillian and the self-energy were computed self-consistently while all vertex corrections were disregarded. Such basic approximations reduce the computational effort considerably but may also limit the range of applicability of the employed methods. We therefore analyzed to what degree such basic approaches take the dominant physical processes reliably into account.

An important test is the benchmark against numerical exact data. In equilibrium, we showed that the RTRG approximation yields reliable results for the position and width of the peaks of the linear conductance that are in very good agreement with highly-accurate NRG data for arbitrary tunneling rates $\Gamma$, despite the fact that the RTRG is perturbative in the coupling between the dot and the reservoirs and is therefore \textit{a priori} controlled only for small tunneling coupling $\Gamma\ll\max\{T,\delta\}$. This means that the charge fluctuations are captured largely by the contribution of the one-loop diagram to the RG equations whereas vertex renormalization seems to be less important to describe these processes. In contrast, cotunneling processes are only partly taken into account, causing deviations of the RTRG solution for the linear conductance from the NRG result in the Coulomb blockade regime, and leading to small anomalies close to the resonances in the case of strong Coulomb interactions. We conclude that the reliability of the RTRG solution depends essentially on the class of diagrams that are resummed and taken into account within the chosen approximation scheme. In this sense, the class of diagrams that is resummed into the renormalized one-loop diagram describes charge fluctuations, while (at least) two-loop diagrams and vertex renormalization are required for a reasonable description of cotunneling processes. The approximation of the RTRG equations can be systematically improved by taking such higher-order diagrams into account, as was already demonstrated in the past for the Kondo model\cite{pletyukhov_schoeller_2012,reininghaus_pletyukhov_schoeller_2014} and the single impurity Anderson model\cite{schoeller_koenig_2000,saptsov_wegewijs_2012}.

In nonequilibrium, we used reliable data for the conductance from the fRG approach in lowest order truncation scheme as a benchmark for the RTRG data for small Coulomb interactions and strong coupling, respectively. Indeed, we find a nearly perfect agreement for the solutions from both approaches in this case, indicating again the drastic extension of the range of validity of the RTRG approximation to arbitrary Coulomb interactions in the regime of charge fluctuations.

We furthermore find from comparing RTRG and fRG solution that the single-particle picture of an \textit{effectively} non-interacting open quantum dot with renormalized parameters is applicable (i) in the regime of small Coulomb interactions $U \ll \Gamma$ and arbitrary bias $V$, and (ii) for intermediate Coulomb interactions that are larger than the bias voltage. This means that the complex interplay between the Coulomb interaction and the tunneling processes away from equilibrium cannot be described by such an effective picture. In agreement with previous RTRG studies of the Anderson model\cite{saptsov_wegewijs_2012}, we showed that the RTRG method is capable of describing this interplay theoretically.

We note that in order to go beyond the effective single-particle picture with the fRG approach, one needs to extend the approximation for the RG equations to the next order. This was demonstrated in the two-level case\cite{jakobs_pletyukhov_schoeller_2010a,karrasch_et_al_2008}, yielding accurate results also for intermediate Coulomb interactions at large bias\cite{eckel_et_al_2010}.

In summary, we advertise the RTRG method as an versatile and flexible tool to describe transport phenomena in quantum dots with an arbitrary geometry in nonequilibrium. In particular, we demonstrated the reliability of this method in describing charge fluctuations in quantum dot systems with a very basic approximation that allows for an efficient numerical computation. We note that the formalism can easily be generalized to finite temperature by calculating the integral in Eq.~(\ref{eq:liouvillian_rg_starting_point}) exactly in terms of the Matsubara poles of the Fermi distribution function. Furthermore, this equation can also be used to calculate the Liouvillian in the whole complex plane for arbitrary $E$ such that the time evolution into the stationary state can be analyzed \cite{schoeller_2014}.

\section*{Acknowledgments}
This work was supported by the Deutsche Forschungsgemeinschaft via RTG 1995 (C.J.L., V.M. and H.S.).
We thank J.~von~Delft, S.~G.~Jakobs, S.-S.~B.~Lee, M.~R.~Wegewijs and A.~Weichselbaum for 
fruitful discussions. F.~B.~K. acknowledges support by the Cluster of 
Excellence Nanosystems Initiative Munich and funding from the research school IMPRS-QST. Numerical calculations for the fRG and RTRG methods were performed with computing resources granted by RWTH Aachen University under project rwth0287.

\begin{appendix}

\section{Perturbation theory for the effective Liouvillian}
\label{app:perturbation_theory}

In this appendix, we discuss bare perturbation theory for the effective Liouvillian and the current kernel of the multi-level Anderson model. The perturbative series can be written as
\begin{align}
\label{eq:liouvillian_series}
L (E) \,&=\, L^{(0)} + L^{(1)} (E) + L^{(2)} (E) + \ldots \quad, \\
\label{eq:current_kernel_series}
\Sigma_\gamma (E) \,&=\, \Sigma_\gamma^{(1)} (E) + \Sigma_\gamma^{(2)} (E) + \ldots \quad,  
\end{align}
where $L^{(m)}(E)$ and $\Sigma^{(m)}_\gamma(E)$, respectively, comprises all diagrams with $m=0,1,2,\ldots$ contraction lines. A contraction represents an excitation in the reservoirs and connects two vertices within a diagram within the diagrammatic language introduced in Refs.~\onlinecite{schoeller_2009,schoeller_2014}. A diagram with $m$ contraction lines is sometimes called a $m$-loop diagram.

The {\it zeroth order} ($m=0$) contribution to the effective Liouvillian is the Liouvillian of the isolated quantum dot, i.e. $L^{(0)} b \,=\,\com{H_\mathrm{s}}{b}$, where $b$ is an arbitrary operator acting on states of the dot Hilbert space. Denoting by $E_s$ the eigenvalues of $H_\mathrm{s}$ and by $\ket{s}$ the corresponding many-body eigenstates, we can express the matrix elements of $L^{(0)}$ as
\begin{align}
\label{eq:liouvillian_zeroth_order}
\lbra{s_1s_2} L^{(0)} \lket{s_1^\prime s_2^\prime} \,=\, \delta_{s_1 s^\prime_1} \delta_{s_2 s^\prime_2} \left(E_{s_1} - E_{s_2}\right) \quad.
\end{align}

Following Refs.~\onlinecite{schoeller_2009,schoeller_2014}, we obtain
\begin{align}
\label{eq:liouvillian_first_order}
L^{(1)}(E) \,=\, \begin{tikzpicture}[baseline=0.8ex,scale=0.5]
\node[node1] (left) at (-1.25,0) {};
\node[node1] (right) at (1.25,0) {};
\draw[line1,black] (left) -- (right) {};
\draw[line1,ForestGreen] (left) -- (-1.25,1) -- (1.25,1) -- (right) {};
\end{tikzpicture} \,=\, L^{(1\text{s})} + L^{(1\text{a})} (E) \quad, 
\end{align}
with
\begin{align}
\notag
L^{(1\text{s})} \,&=\, \int \mathrm{d} \omega\, \gamma^\mathrm{s}_{11^\prime} (\omega) G_1 \frac{1}{E + \omega + \overline{\mu}_\alpha - L^{(0)}} \tilde{G}_{1^\prime} \\
\label{eq:liouvillian_symmetric_part}
&=\, - i \frac{\pi}{2} G_1 \tilde{G}_{\overline{1}} \quad, \\
\notag
L^{(1\text{a})} (E) \,&=\, \int \mathrm{d} \omega\, \gamma^\mathrm{a}_{11^\prime} (\omega) G_1 \frac{1}{E + \omega + \overline{\mu}_\alpha - L^{(0)}} G_{1^\prime} \\
\label{eq:liouvillian_antisymmetric_part}
&=\, G_1 \ln \frac{-i\left(E + \overline{\mu}_\alpha - L^{(0)} \right)}{D} G_{\overline{1}} \quad,
\end{align}
for the \textit{first order} correction to the effective Liouvillian. The leading order term for the current kernel can be obtained from these equations by simply replacing the left vertex $G_1$ by the current vertex \eqref{eq:current_vertex} in all expressions, yielding
\begin{align}
\label{eq:current_kernel_symmetric_part}
\Sigma^{(1\text{s})}_{\gamma} \,&=\, - i \frac{\pi}{2} c^\gamma_1 \tilde{G} _1 \tilde{G}_{\overline{1}} \quad, \\
\label{eq:current_kernel_antisymmetric_part}
\Sigma^{(1\text{a})}_{\gamma} (E) \,&=\, c^\gamma_1 \tilde{G} _1 \ln \frac{-i\left(E + \overline{\mu}_\alpha - L^{(0)} \right)}{D} G_{\overline{1}} \quad.
\end{align}

In the first lines of equations \eqref{eq:liouvillian_symmetric_part} and \eqref{eq:liouvillian_antisymmetric_part},
\begin{align}
\label{eq:contraction_decomposition}
\gamma^{\mathrm{s},\mathrm{a}}_{11^\prime} (\omega) \,&=\, \delta_{\eta,-\eta^\prime} \delta_{\alpha\alpha^\prime} \rho_\mathrm{c} (\omega) f^{\mathrm{s},\mathrm{a}}_{\alpha} (\omega) \quad,
\end{align}
are the symmetric and antisymmetric part of the contraction $\gamma^{pp^\prime}_{11^\prime} (\omega) \,=\, p^\prime \gamma^\mathrm{s}_{11^\prime} (\omega) + \gamma^\mathrm{a}_{11^\prime} (\omega)$. Accordingly, $f^{\mathrm{s},\mathrm{a}}_{\alpha} (\omega) = {1 \over 2} [f(\omega) \pm f(-\omega)]$ are the symmetric and antisymmetric part of the Fermi distribution. The former always gives $f^{\mathrm{s}}_{\alpha} (\omega)\,=\, {1 \over 2}$ while the latter $f^{\mathrm{a}}_{\alpha} (\omega)\,=\, -{1 \over 2} \sgn(\omega)$ for $T_\alpha=0$. Furthermore, we have incorporated the factor $p^\prime$ in front of $\gamma^\mathrm{s}_{11^\prime} (\omega)$ into the second vertex in \eqref{eq:liouvillian_symmetric_part} and \eqref{eq:current_kernel_symmetric_part}, yielding $\tilde{G}_1 \,=\, \sum_{p=\pm} p G^p_1$.

We have introduced the Lorentzian high-frequency cut-off $\rho_\mathrm{c} (\omega)=D^2/(\omega^2 + D^2)$ with bandwidth $D \rightarrow \infty$ in order to regularize the frequency integral for high frequencies which results in the term $\sim \ln D$ in \eqref{eq:liouvillian_antisymmetric_part}. However, this term drops out since 
\begin{align}
\notag
G_1 G_{\overline{1}} \,&=\, \sum_{pp^\prime} \sum_{\eta l_1 l_2} t^\eta_{\alpha l l_1} t^{-\eta}_{\alpha l l_2} C^p_{\eta l_1} C^{p^\prime}_{-\eta l_2} \\
\notag
&=\, \frac{1}{2} \sum_{pp^\prime} \sum_{\eta l_1 l_2} t^\eta_{\alpha l l_1} t^{-\eta}_{\alpha l l_2} \anticom{C^p_{\eta l_1}}{C^{p^\prime}_{-\eta l_2}} \\
\notag
&=\,  \frac{1}{2} \sum_{p} \sum_{l_1} t^\eta_{\alpha l l_1} t^{-\eta}_{\alpha l l_1} p \\
\label{eq:special_relation_1}
&=\,0
\end{align} 
where we used the anticommutation relation $\anticom{C_{\eta l}^{p}}{C_{\eta^\prime l^\prime}^{p^\prime} } \,=\, p \delta_{p p^\prime} \delta_{\eta, - \eta^\prime} \delta_{l l^\prime}$ for the dot field superoperators after the second line. In order to show that the term $\sim \ln D$ in the last line in \eqref{eq:current_kernel_antisymmetric_part} can be disregarded similarly, we note that we only need the combination $\tr_\mathrm{s} \Sigma_\gamma (E)$ in order to compute the current $I_\gamma$ from \eqref{eq:current_rtrg}. From the general property\cite{schoeller_2009,schoeller_2014} $\tr_\mathrm{s} G_1^p =0$ one can deduce
\begin{align}
\label{eq:current_vertex_trace}
\tr_\mathrm{s} \tilde{G}_1 \,=\, 2 \tr_\mathrm{s} G^+_1 \,=\, - 2 \tr_\mathrm{s} G^-_1 \,=\, - 2p^\prime \tr_\mathrm{s} G^{-p^\prime}_1 \quad,
\end{align}
which leads to
\begin{align}
\notag
\tr_\mathrm{s} c^\gamma_1 \tilde{G}_1 G_{\overline{1}} \,&=\, - 2\tr_\mathrm{s} \sum_{p^\prime} \sum_{\eta l_1 l_2} \eta p^\prime t^\eta_{\alpha l l_1} t^{-\eta}_{\alpha l l_2} C^{-p^\prime}_{\eta l_1} C^{p^\prime}_{-\eta l_2} \\
\notag
&=\, - \tr_\mathrm{s} \sum_{p^\prime} \sum_{\eta l_1l_2} \eta p^\prime t^\eta_{\alpha l l_1} t^{-\eta}_{\alpha l l_2} \anticom{C^{-p^\prime}_{\eta l_1}}{ C^{p^\prime}_{-\eta l_2}} \\
\label{eq:special_relation_2}
&=\, 0 \quad.
\end{align}   
Thus, we can equivalently consider
\begin{align}
\label{eq:liouvillian_antisymmetric_part_no_scale}
L^{(1\text{a})} (E) \,&=\, G_1 \ln -i\left(E + \overline{\mu}_\alpha - L^{(0)} \right) G_{\overline{1}} \quad,\\
\label{eq:current_kernel_antisymmetric_part_no_scale}
\Sigma^{(1\text{a})}_{\gamma} (E) \,&=\, c^\gamma_1 \tilde{G} _1 \ln -i\left(E + \overline{\mu}_\alpha - L^{(0)} \right) G_{\overline{1}} \quad,
\end{align}
instead of \eqref{eq:liouvillian_antisymmetric_part} and \eqref{eq:current_kernel_antisymmetric_part}. Importantly, \eqref{eq:special_relation_1} and \eqref{eq:special_relation_2} have the consequence that perturbation theory yields no logarithmic divergences in the ultraviolet regime $\abs{E} \rightarrow \infty$. A resummation of logarithmic terms is therefore not necessary in this case. This explains why we can neglect vertex corrections in lowest order truncation for the RG treatment. Thus, we can simply insert the bare vertices $G_1$ and $(I_\gamma)_1$ into the RG equations.

In particular, the only logarithmic singularities of the effective Liouvillian and the current kernel for $E=i0^+$ are given by the condition \eqref{eq:resonant_tunneling_pt}. In order to treat these singularities, it is sufficient to calculate the effective Liouvillian self-consistently, which is achieved by the RTRG approach. The consequence is that the complex eigenvalues $\lambda_k(E)$ of the effective Liouvillian and {\it not} the real eigenvalues $E_{s_1} - E_{s_2}$ of the bare Liouvillian $L^{(0)}$ enter the argument of the complex logarithm in \eqref{eq:liouvillian_antisymmetric_part_no_scale} and \eqref{eq:current_kernel_antisymmetric_part_no_scale}. The imaginary part of $\lambda_k(E)$ provide a cut-off that regularizes the logarithms. The sole exception is the non-degenerate eigenvalue $\lambda_\text{st}=0$ which, however, never appears in the argument of the logarithm, as discussed in more detail in Refs.~\onlinecite{schoeller_2009,schoeller_2014}.

\textit{Second order} diagrams ($m=2$) are necessary to describe \textit{cotunneling processes}. The two contraction lines in these diagrams account for the two excitations generated in the reservoir in a flavor fluctuation due to the coupling between dot and reservoir. One finds that the second order contribution is given by the two diagrams
\begin{align*}
&\begin{tikzpicture}[baseline=0.8ex,scale=0.5]
\node[node1] (left1) at (-3.75,0) {};
\node[node1] (left2) at (-1.25,0) {};
\node[node1] (right1) at (1.25,0) {};
\node[node1] (right2) at (3.75,0) {};
\draw[line1,black] (left1) -- (left2) {};
\draw[line1,black] (left2) -- (right1) {};
\draw[line1,black] (right1) -- (right2) {};
\draw[line1,ForestGreen] (left1) -- (-3.75,1.4) -- (3.75,1.4) -- (right2) {};
\draw[line1,ForestGreen] (left2) -- (-1.25,1) -- (1.25,1) -- (right1) {};
\end{tikzpicture} 
\quad,\\
&\begin{tikzpicture}[baseline=0.8ex,scale=0.5]
\node[node1] (left1) at (-3.75,0) {};
\node[node1] (left2) at (-1.25,0) {};
\node[node1] (right1) at (1.25,0) {};
\node[node1] (right2) at (3.75,0) {};
\draw[line1,black] (left1) -- (left2) {};
\draw[line1,black] (left2) -- (right1) {};
\draw[line1,black] (right1) -- (right2) {};
\draw[line1,ForestGreen] (left1) -- (-3.75,1) -- (1.25,1) -- (right1) {};
\draw[line1,ForestGreen] (left2) -- (-1.25,1.4) -- (3.75,1.4) -- (right2) {};
\end{tikzpicture} \quad.
\end{align*}
The upper diagram contains a connected first-order subdiagram as insertion on the propagator line. It belongs to the class of connected subdiagrams with no free contraction lines, i.e. all contraction lines connect two vertices of this subdiagram. These subdiagrams are sometimes called \textit{self-energy insertion}, although they have nothing to do with the physical self-energy of a single-particle Green's function, apart from a formal equivalence. Resumming these insertions, one can replace all free propagators by full ones which leads to self-consistent perturbation theory\cite{schoeller_2014}. Since the diagram on the right-hand side of the RG equation \eqref{eq:liouvillian_rg_starting_point} contains only the full propagator, the upper diagram is also included in the RTRG approximation discussed in section \ref{subsec:rtrg}. In contrast, the diagram with the crossed contraction lines are not included within the considered truncation scheme. To include also this diagram, one needs to add the corresponding two-loop diagram on the right hand side of the RG equation \eqref{eq:liouvillian_rg_starting_point} as well as to include the vertex correction by replacing the bare vertex by the effective one. The latter can then be obtained as solution of a corresponding RG equation.

Finally, we note that there are also no logarithmic divergent terms in the ultraviolet limit $\abs{E} \rightarrow \infty$ in higher-order perturbation theory. A $m$-th order diagram consisting of $m$ contraction lines and $2m$ vertices contains $2m-1$ resolvents $\sim (E_{1\ldots n} + \underline{\omega}_{1 \ldots n} - L^{(0)})^{-1}$ with $E_{1\ldots n}\,=\, E + \sum_{k=1}^n \overline{\mu}_k$ and $\overline{\omega}_{1 \ldots n} \sum_{k=1}^n \overline{\omega}_k$. Since each contraction gives rise to one frequency integral, one can estimate that the $m$-th order diagram with $m \geq 2$ falls off $\sim E^{1-m}$ for $\abs{E} \rightarrow \infty$.

In the same way, all $m$-th order diagrams with $m \geq 2$ vanish in the limit $\abs{\overline{\mu}_\alpha} \rightarrow \infty$. In the case $m=1$, we find that the part of the diagram with the antisymmetric part of the contraction $\gamma_{11^\prime}^\mathrm{a}(\omega)$ vanishes for $\abs{\overline{\mu}_\alpha} \rightarrow \infty$ due to the property \eqref{eq:special_relation_1}. As a consequence, the effective Liouvillian is given by \eqref{eq:liouvillian_infinite_bias} in this case.

\section{Truncation of the RTRG equation}
\label{app:truncation}

After equation \eqref{eq:current_kernel_rg_real_flow_parameter}, we have explained  that \eqref{eq:liouvillian_rg_real_flow_parameter} defines effectively an infinite hierarchy of RG equations. In order to truncate this hierarchy of RG equations, we bring this system in a more transparent form for the special case of two reservoirs. Following Ref.~\onlinecite{reininghaus_pletyukhov_schoeller_2014}, we define a chain of discrete points
\begin{align}
\label{eq:bias_chain_points}
\mu_k \,=\, \frac{k}{2} V \quad,
\end{align}
with an integer number $k$. Obviously, $k=1$ and $k=-1$ correspond to the chemical potentials of the two reservoirs, i.e. $\mu_1\,=\,\mu_L$ and $\mu_{-1}\,=\,\mu_R$, respectively. With the definition
\begin{align}
\label{eq:bias_chain_liouvillian}
\tilde{L}_k (\Lambda) \,=\, \tilde{L} (\Lambda - i \mu_k) \quad,
\end{align}
the aforementioned hierarchy of RG equations is given by
\begin{widetext}
\begin{align}
\label{eq:liouvillian_rg_equation_bias_chain}
\td{}{\Lambda} \tilde{L}_k (\Lambda) \,=\, i \sum_{\eta \alpha l} G_{\eta \alpha l}\, \frac{1}{i \Lambda + \mu_{k + \nu_{\alpha \eta}} - \tilde{L}_{k + \nu_{\alpha \eta}}(\Lambda)} \, G_{-\eta \alpha l} \quad,
\end{align}
where we have introduced the sign factor
\begin{align}
\label{eq:sign_factor}
\nu_{\alpha \eta} \,=\, \begin{cases} 
+1 & \text{if} \quad \eta=+,\, \alpha=\text{L} \quad \text{or} \quad \eta=-,\, \alpha=\text{R} \\
-1 & \text{if} \quad \eta=+,\, \alpha=\text{R} \quad \text{or} \quad \eta=-,\, \alpha=\text{L}
\end{cases} \quad.
\end{align}
Within this notation scheme, the RG equation for the current kernel \eqref{eq:current_kernel_rg_real_flow_parameter} recast as
\begin{align}
\label{eq:current_kernel_rg_equation_bias_chain}
\td{}{\Lambda} \tilde{\Sigma}_\alpha (\Lambda) \,&=\, - \frac{i}{2} \sum_{l \eta} \eta\, \tilde{G}_{\eta \alpha l} \frac{1}{i \Lambda + \mu_{\nu_{\alpha \eta}} - \tilde{L}_{\nu_{\alpha \eta}}(\Lambda)} G_{-\eta \alpha l} \quad.
\end{align} 
\end{widetext}
The initial conditions are
\begin{align}
\label{eq:liouvillian_initial_bias_chain}
\tilde{L}_k(\Lambda) \big|_{\Lambda=D} \,&=\, L^{(0)} + L^{(1\mathrm{s})} \quad,
\end{align}
since \eqref{eq:liouvillian_initial} holds for any $k$.

Truncation of the infinite hierarchy of RG equations is achieved by setting
\begin{align}
\label{eq:liouvillian_truncation}
\tilde{L}_{\pm (k_0 + 1)}(\Lambda)\,\approx\, \tilde{L}_{\pm k_0}(\Lambda)
\end{align}
for some $k_0$. This is justified due to
\begin{align}
\label{eq:energy_shift_relative_change}
\frac{\mu_{k+1} - \mu_k}{\mu_k} \,=\, \frac{1}{k} \quad,
\end{align}
which means that the relative change in the energy shift $\mu_k$ in the argument of the Liouvillian $\tilde{L}(\Lambda - i \mu_k)$ falls off $\sim k^{-1}$ for $k \rightarrow \infty$. In practice, we have checked convergence of the solution by comparing the results for different values of $\abs{k_0}$. We consider a solution as reliable if the result for this choice does not deviate significantly from the one obtained for $\abs{\tilde{k}_0}\,=\,\abs{k_0}+1$. For all numerical calculations, we observed a convergence already for quite small values of $\abs{k_0}$. In particular, $\abs{k_0}\,=\,4$ proved to be a reliable choice for all cases considered in this article.

\begin{widetext}
\section{Closed analytic expressions of the fRG equation for the self-energy and the current}
\label{app:selfenergy}

The integral on the right hand side of \eqref{eq:retarted_selfenergy_rq_equation_starting_point} can be analytically evaluated, as we discuss now. Inserting \eqref{eq:keldysh_singlescale_propagator} into \eqref{eq:retarted_selfenergy_rq_equation_starting_point} gives
\begin{align}
\label{eq:retarded_selfenergy_rg_equation_alternative_form}
\td{}{\Lambda} \Sigma^\text{R}_{ll^\prime}(\Lambda) \,=\, - \frac{1}{4 \pi} \sum_{l_1 l_1^\prime} \overline{v}_{l l_1, l^\prime l^\prime_1}  \int \mathrm{d} \omega \, \left[\dunderline{G}^\text{R}(\Lambda,\omega) \dunderline{G}^\text{K}(\Lambda,\omega)  -  \dunderline{G}^\text{K}(\Lambda,\omega) \dunderline{G}^\text{A}(\Lambda,\omega) \right]_{l^\prime_1 l_1} \quad.
\end{align}
To evaluate the integral, we furthermore make use of the spectral representation of the retarded and advanced, respectively, component of the self-energy, i.e.
\begin{align}
\label{eq:retarded_selfenergy_spectral_decomposition}
\dunderline{\Sigma}^{\text{R}} (\Lambda) \,&=\, \sum_k \lambda_k^\Lambda \dunderline{P}_k^\Lambda \quad, \\
\label{eq:advanced_selfenergy_spectral_decomposition}
\dunderline{\Sigma}^{\text{A}} (\Lambda) \,&=\, \sum_k \left(\lambda_k^\Lambda \right)^{*} \left( \dunderline{P}_k^\Lambda \right)^\dagger \quad.
\end{align}
Inserting \eqref{eq:retarded_greens_function}, \eqref{eq:keldysh_greens_function}, \eqref{eq:retarded_selfenergy_spectral_decomposition} and \eqref{eq:advanced_selfenergy_spectral_decomposition} into \eqref{eq:retarded_selfenergy_rg_equation_alternative_form} and using the integral
\begin{align}
\label{eq:integral_3}
\int \mathrm{d}\omega\,\sgn(\omega)\, \frac{1}{(\omega + z_1)^2} \frac{1}{\omega + z_2} \,=\,\frac{2}{z_1 - z_2} \left\lbrace \frac{1}{z_1 - z_2} \left[ \ln (-i \sigma_1 z_1) - \ln (-i \sigma_2 z_2) \right] - \frac{1}{z_1} \right\rbrace
\end{align}
with $\sigma_i \,=\, \sgn (\imag z_i)$ yields
\begin{align}
\notag
\td{}{\Lambda} \Sigma_{ll^\prime}^\text{R}(\Lambda) \,&=\, \frac{i}{2 \pi} \sum_{l_1 l_1^\prime} \overline{v}_{l l_1, l^\prime l^\prime_1} \sum_{\alpha kk^\prime} \left[ \dunderline{P}_k^\Lambda \dunderline{\Gamma}^\alpha \left(\dunderline{P}_{k^\prime}^\Lambda \right)^\dagger \right]_{l^\prime_1 l_1} \frac{1}{\lambda_k^\Lambda - \left(\lambda_{k^\prime}^\Lambda \right)^* - 2 i \Lambda} \\
\notag
& \quad \times \left\lbrace \frac{1}{\mu_\alpha - \lambda_k^\Lambda + i \Lambda} + \frac{1}{\mu_\alpha - \left(\lambda_{k^\prime}^\Lambda \right)^* -  i \Lambda} \right. +  \frac{2}{\lambda_k^\Lambda - \left(\lambda_{k^\prime}^\Lambda \right)^* - 2 i \Lambda} \\
\label{eq:retarded_selfenergy_rg_equation}
& \quad \times \left. \left[ \ln -i(\mu_\alpha - \lambda_k^\Lambda + i \Lambda)  - \ln i(\mu_\alpha - \left(\lambda_{k^\prime}^\Lambda \right)^* -  i \Lambda) \right] \right\rbrace \quad.
\end{align}
In the same way, we can evaluate the frequency integral in the current formula \eqref{eq:current_frg}. Using the results
\begin{align}
\label{eq:integral_1}
\int \mathrm{d}\omega\,\sgn(\omega)\frac{1}{\omega + z_1} \frac{D^2}{D^2 + \omega^2} \,&=\, - 2 \frac{D^2}{D^2 + z_1^2} \ln \frac{-i \sigma_1 z_1}{D} \xrightarrow{D \rightarrow \infty}\, - 2 \ln \frac{-i \sigma_1 z_1}{D} \quad,\\ 
\label{eq:integral_2}
\int \mathrm{d}\omega\,\sgn(\omega)\, \frac{1}{\omega + z_1} \frac{1}{\omega + z_2} \,&=\, \frac{2}{z_1 - z_2} \left[ \ln( -i \sigma_1 z_1) - \ln (-i \sigma_2 z_2) \right]\quad,
\end{align}
we obtain
\begin{align}
\notag
I^\text{st}_\alpha \,&=\, \frac{i}{2 \pi} \sum_k \tr \left\lbrace \dunderline{\Gamma}^\alpha \left[ \ln -i(\mu_\alpha - \tilde{\lambda}_k) \tilde{\dunderline{P}}_k - \ln i(\mu_\alpha - \tilde{\lambda}_k^*) \tilde{\dunderline{P}}_k^\dagger \right] \right\rbrace \\
\notag
&\quad - \frac{1}{2 \pi} \sum_{\alpha^\prime k k^\prime} \frac{1}{\tilde{\lambda}_k - \tilde{\lambda}_{k^\prime}^*} \left[ \ln -i(\mu_{\alpha^\prime} - \tilde{\lambda}_k)  - \ln i(\mu_{\alpha^\prime} - \tilde{\lambda}_{k^\prime}^*) \right] \tr \left( \tilde{\dunderline{P}}_k\, \dunderline{\Gamma}^{\alpha^\prime}\, \tilde{\dunderline{P}}_{k^\prime}^\dagger\, \dunderline{\Gamma}^\alpha \right) \\
\notag
&=\, \frac{1}{2 \pi} \real \tr \left\lbrace 2i \sum_k \ln -i(\mu_\alpha - \tilde{\lambda}_k) \tilde{\dunderline{P}}_k \, \dunderline{\Gamma}^\alpha - \sum_{\alpha^\prime k k^\prime} \tilde{\dunderline{P}}_k \, \dunderline{\Gamma}^{\alpha^\prime} \, \tilde{\dunderline{P}}_{k^\prime}^\dagger \, \dunderline{\Gamma}^\alpha \right. \\
\label{eq:current_frg_spectral_representation}
&\quad \left. \times \frac{1}{\tilde{\lambda}_k - \tilde{\lambda}_{k^\prime}^*} \left[ \ln -i(\mu_{\alpha^\prime} - \tilde{\lambda}_k)  - \ln i(\mu_{\alpha^\prime} - \tilde{\lambda}_{k^\prime}^*) \right] \right\rbrace\quad,
\end{align}
with $\tilde{\lambda}_k\,=\, \lambda_k^{\Lambda=0}$ and $\tilde{\dunderline{P}}_k\,=\,\dunderline{P}_k^{\Lambda=0}$.
\end{widetext}

\end{appendix}

\end{document}